%% file: paper.tex
\documentclass{article}

\usepackage{parskip}
\usepackage{amsmath, amssymb}
\usepackage[table]{xcolor}
\usepackage{graphicx}
\usepackage[numbers]{natbib}
\usepackage{tikz}
\usepackage{subcaption}
\usepackage{hyperref}
\usepackage[left=2cm, right=2cm]{geometry}
\usepackage{booktabs}
\usepackage{float}

\allowdisplaybreaks

\title{Multivariate Simulation-based Forecasting for Intraday Power Markets: Modelling Cross-Product Price Effects}
\author{%
        Simon Hirsch$^{1, 2}$ \\ 
        \texttt{\href{mailto:simon.hirsch@stud.uni-due.de}{simon.hirsch@stud.uni-due.de}}\\ 
        \texttt{\href{mailto:simon.hirsch@statkraft.com }{simon.hirsch@statkraft.com }}
    \and 
        Florian Ziel$^1$ \\ 
        \texttt{\href{mailto:florian.ziel@uni-due.de}{florian.ziel@uni-due.de} }
}
\date{%
    $^1$House of Energy Markets and Finance, University of Duisburg-Essen\\%
    $^2$Statkraft Trading GmbH\\[2ex]%
    \today
}

\begin{document}
\maketitle
\begin{abstract}
    Intraday electricity markets play an increasingly important role in balancing the intermittent generation of renewable energy resources, which creates a need for accurate probabilistic price forecasts. However, research to date has focused on univariate approaches, while in many European intraday electricity markets all delivery periods are traded in parallel. Thus, the dependency structure between different traded products and the corresponding cross-product effects cannot be ignored. We aim to fill this gap in the literature by using copulas to model the high-dimensional intraday price return vector. We model the marginal distribution as a zero-inflated Johnson's $S_U$ distribution with location, scale and shape parameters that depend on market and fundamental data. The dependence structure is modelled using latent beta regression to account for the particular market structure of the intraday electricity market, such as overlapping but independent trading sessions for different delivery days. We allow the dependence parameter to be time-varying. We validate our approach in a simulation study for the German intraday electricity market and find that modelling the dependence structure improves the forecasting performance. Additionally, we shed light on the impact of the single intraday coupling (SIDC) on the trading activity and price distribution and interpret our results in light of the market efficiency hypothesis. The approach is directly applicable to other European electricity markets.
\end{abstract}
\textit{Keywords:} Intraday Electricity Markets, Electricity Price Forecasting, Volatility Forecasting, Copula, Probabilistic Forecasting, Monte-Carlo Methods, SIDC \vspace{0.5cm} \newline 
\textit{Acknowledgements:} Simon Hirsch is employed at Statkraft Trading GmbH and gratefully acknowledges support through Statkraft (\url{https://www.statkraft.com/}). This work contains the author’s opinion and does not necessarily reflect Statkraft’s position. The authors declare no conflict of interest. \newline

\section{Introduction}

Intraday electricity markets are used to balance the short-term intermittency of renewable energy assets. The increasing penetration of wind and solar yields the need for accurate probabilistic price forecasts. This need is underscored by the heightened volatility in light of the European energy crisis in 2022/23. To date, the literature on (probabilistic) electricity price forecasting on intraday markets remains scarce \cite{narajewski2020ensemble, hirsch2022simulation, uniejewski2019understanding, narajewski2020econometric} and is exclusively focused on univariate approaches. However, in most European countries, the intraday market is a continuous forward market, where all delivery periods for a delivery day $d$ are traded in parallel, thus univariate approaches neglect the complex dependency structure in these markets. As products close with the physical delivery of electricity, it is not possible to ``glue'' subsequent trading sessions together, as it is commonly done with equity markets. Additionally, while the spot market is driven by the absolute level of fundamentals such as wind and solar production, the intraday prices are influenced more by the changes in forecasts \cite{ziel2017modeling}. Our work focuses on these challenges in modelling the dependency structure and fills the according gap in the literature. Based on the work of \cite{narajewski2020ensemble, hirsch2022simulation} on the marginal distribution of the price process, we employ copulas to model the time-dependent correlation. We model the dependency parameter as latent variable using beta regression. We validate our results in a forecasting study for the German intraday electricity market. Our results indicate that modelling the dependence structure between the different trading session improves the forecasting performance. Additionally, we provide evidence on market efficiency in the German short-term market. Our fundamental and parametric approach allows us to shed further light on the impact of the cross-border shared order books of the single intraday coupling (SIDC) and the driving factors of the distribution parameters.  

\begin{figure}
    \centering
    \includegraphics[width=\textwidth]{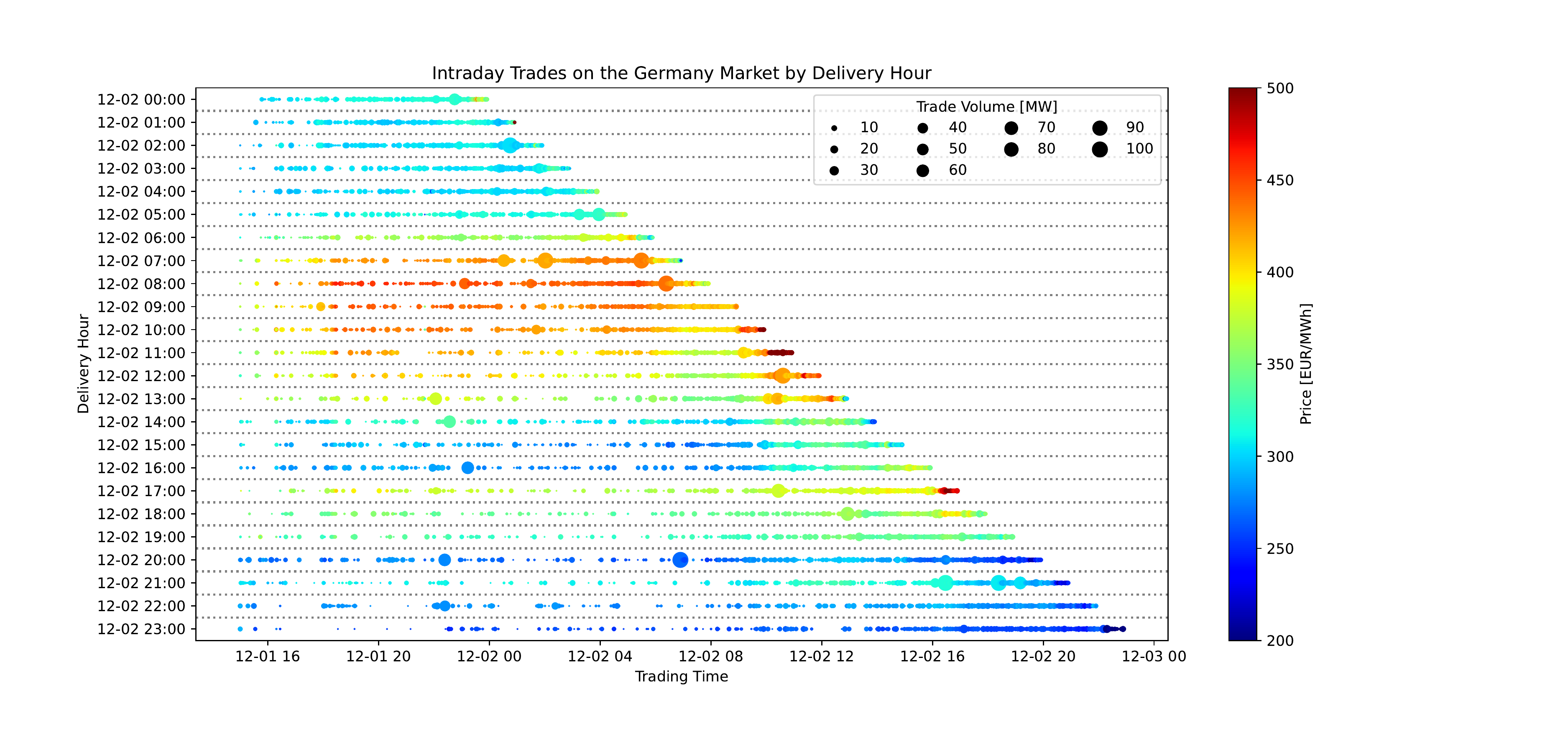}
    \caption{All trades for delivery on 2022-12-02 on the German continuous intraday market by delivery period and trading time. Trading for all delivery periods opens on 2022-12-01 at 15:00 hours. Each dot represents one trade. The size of the dots corresponds to the size of individual trades in MW. The color reflects the transacted price in EUR/MWh. The correlated behaviour of neighbouring trading periods is clearly visible.}
    \label{fig:trading_in_parallel_trading_sessions}
\end{figure}

Let us give an illustrative example for the trading schedule of the German intraday electricity market. Trading for physical delivery on $d$ starts on the previous day $d-1$, 15:00 hours and lasts till few minutes before the start of physical delivery of electricity. For example, trading for the delivery on $d$, 18:00-19:00 started on $d-1$ at 15:00 hours and closes at $d$, 17:55 hours. During the beginning of this trading window, traders will be able to trade power in many neighbouring delivery hours, from delivery at $d$ 16:00-17:00 (which closes at $d-1$ 15:55), to all delivery periods for the next delivery day $d+1$ (which start trading at $d$, 15:00 hours). Figure \ref{fig:trading_in_parallel_trading_sessions} shows all trades in the two trading sessions for delivery on 2022-12-01 and 2022-12-02 for all hourly products on the German intraday market. 

The literature on intraday electricity price forecasting can be divided into three main groups: $(1)$ paper which treat the intraday market in a similar fashion as the day-ahead market and predict (index) prices along the delivery time line \cite{uniejewski2019understanding, cramer2022multivariate, kath2021conformal} and $(2)$ paper which predict prices along the trading time for single delivery periods \cite{serafin2022trading, hirsch2022simulation, narajewski2020econometric, narajewski2020ensemble, janke2019forecasting, marcjasz2020beating}. The correlation between different trading windows has, to the best knowledge of the authors, not been investigated so far, while the correlation between day-ahead and intraday (index) prices has been the subject of studies such as \cite{andrade2017probabilistic, cramer2022multivariate}. Lastly, $(3)$, empirical, in-sample studies on price formation in intraday electricity markets have been conducted by \cite{kiesel2017econometric, kremer2021econometric, kath2019modeling}.
 
In light of the aforementioned challenges and the gap identified in the literature, our contributions are:
\begin{itemize}
    \item We develop a global model for the marginal distribution of intraday electricity prices in the German market and extend previous work from \cite{narajewski2020ensemble, hirsch2022simulation} by taking the whole trading window into account. 
    \item As novelty, we analyse the correlation and dependence structure in the German intraday electricity market and develop a multivariate, probabilistic forecasting model for the German intraday electricity markets to take cross-product effects into account.
    \item We validate our approach in an extensive simulation study for the German intraday electricity market.
    \item Using a parametric approach, our methods and models shed light on the driving variables such as trading activity and renewable forecasts in the intraday market, but also on the impact of the market structure and SIDC. 
\end{itemize}

\begin{figure}
    \centering
    \includegraphics[width=\textwidth]{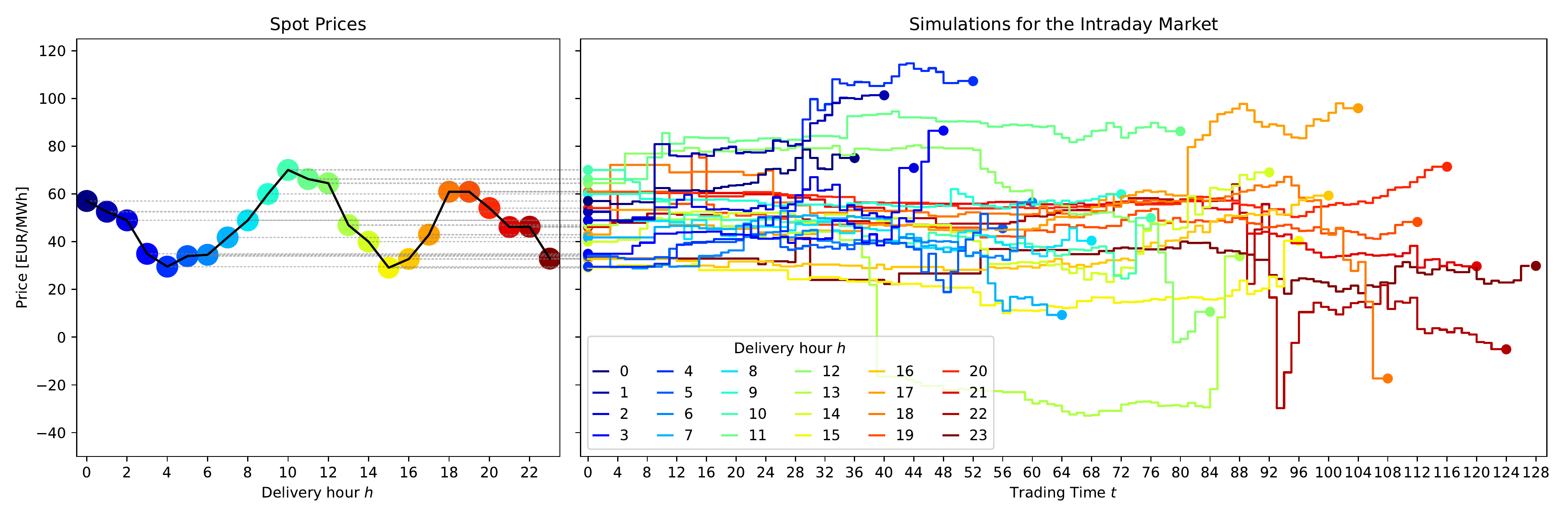}
    \caption{One exemplary simulation for all delivery hours of 2022-01-02. All simulations start at the day-ahead spot price and develop correlated along the trading time $t$.}
    \label{fig:example_simulations}
\end{figure}

Our main strategy is the canonical inference-for-margins \cite{patton2012review} approach commonly used for copula modelling and can be summarized as follows: We use the probabilistic models developed by \cite{narajewski2020ensemble, hirsch2022simulation} as a starting point to \emph{gaussianize} the intraday market observations. We use the pseudo-Gaussian observations to fit the dependency structure. For forecasting, we simulate (multivariate) Gaussian random variates and use the inverse probability integral transformation to receive samples in the desired marginal distribution. Within the energy markets literature, similar approaches have been used for simulating wind and load forecast deviations \cite{tastu2015space, carmona2021glasso, carmona2022joint}, day-ahead electricity prices \cite[see e.g.][]{manner2019forecasting, pircalabu2017regime, berrisch2023modeling} and the design of hedging strategies \cite{pircalabu2017mixed}. Our results show that modelling the dependence structure leads to significantly improved forecasting performance compared to univariate approaches. However, we find that time-dependent modelling of the dependence structure is of little added value compared to constant dependence. 
Additionally, we provide new insight on the effects of the opening and closing of the cross-border order books during SIDC. We interpret our results in light of the market efficiency hypothesis and discuss reflections on modelling already highly volatile prices during a period of increased uncertainty. Figure \ref{fig:example_simulations} gives an illustrative example of our 24-dimensional forecast

Our results offer multiple avenues for further research. First, we restrict ourselves to Gaussian dependence structures. Future work might improve our methods by using copulas that reflect possible tail dependence effects or use Vine-copulas to better approximate the dependence structure. Secondly, we note that the zero-inflated Johnson's $S_U$ distribution offers a good, but not yet perfect fit for the marginal distribution in the intraday market and hence further research in the marginal distribution of intraday electricity prices is required. Third, in light of the  continued development of SIDC and the planned introduction of intraday auctions within the SIDC system and the possible integration of interconnector cables in the SIDC system \cite{entsoe2023single}, our results on the impact of SIDC on the trading activity provide a fruitful starting point for further research \cite[for an early work on the topic see][]{kath2019modeling}.

What is more, our results are also relevant for researchers and practitioners working on stochastic optimization of bidding strategies for the intraday markets. For storage assets such as batteries and pumped-hydro, modelling the dependency structure is important as charging (pumping) positions some periods depends on the ability to discharge (generate) later during the day and therefore depends on the dependency structure. Recent works as \cite{boukas2021deep, nolzen2022market, finhold2023optimizing} use sampled paths for the intraday market, but model the dependency structure only implicitly, if at all, and thus might produce too optimistic results. 

The remainder of this paper is structured as follows: the following Section \ref{sec:market} gives a detailed introduction to the German short-term electricity market. Section \ref{sec:data} introduces our data set, preparation and summary statistics. We present our modelling approach in Section \ref{sec:models}. Section \ref{sec:study_design} describes the forecasting study design and scoring rules. Finally, Section \ref{sec:results} and \ref{sec:discussion} present and discuss our results and conclude this paper. 

\section{Market Description} \label{sec:market}

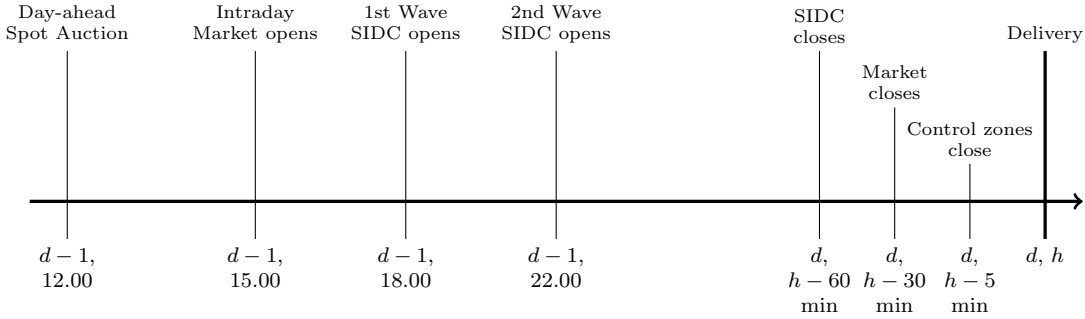
\begin{figure}[htb]
\begin{center}
\input{figures/fig_market_structure_shortterm.tex}
\caption{Daily procedure in the German short-term power markets \citep[based on][]{epex2021trading}. Note that the figure abstracts from half-hourly and quarter-hourly delivery periods.} 
\label{fig:market_structure_shortterm_sequential}
\end{center}
\end{figure}

Electricity markets are structured as forward markets. The German short-term electricity market consists of three major parts: (1) the daily spot auction, (2) the continuous intraday market and (3) the balancing market. The following discussion focuses on the spot and intraday markets. The spot market is the main electricity market in Germany. It is organized as a daily auction at noon on which electricity for all 24 delivery hours on the following day is traded. The following intraday market is used to balance deviations in forecasts after the day-ahead market. It is organized as continuous trading similar to equity markets. The continuous trading starts at $d-1$, 15:00 hours and closes shortly before delivery. After the delivery period ends, remaining imbalances between the traded position and the actual production are settled in the balancing market with the TSO. However, strict market regulation in Germany prohibit explicit active position taking in the balancing market. Figure \ref{fig:market_structure_shortterm_sequential} depicts the daily procedure for a single delivery hour. 

Let us introduce some nomenclature to ease the following discussion of the market structure of the spot and intraday market. We refer to the \emph{delivery time} as the time of actual production power, while the \emph{trading time} refers to the time at which a trade for a certain delivery period is conducted. As a general rule, we try to denote delivery time in superscript, while we denote trading time in subscript. We refer to a \emph{trading session} on the intraday market as the time window between market opening on $d-1$, 15:00 hours to gate closure. Note that for different delivery periods, trading sessions are of different length.

The spot market in Germany is organized by EPEX Spot and Nordpool AS with shared order books. On the spot market, electricity for all 24 hours for the following day is traded. The market is organized as a pay-as-cleared auction and its order book closes at $d-1$, 12:00 hours. Results are published at $d-1$, 12:42 hours. The minimum price is currently set to -500 EUR/MWh and the maximum price is set to 3000 EUR/MWh. 

\begin{figure}[htb]
    \centering
    \includegraphics[width=\textwidth]{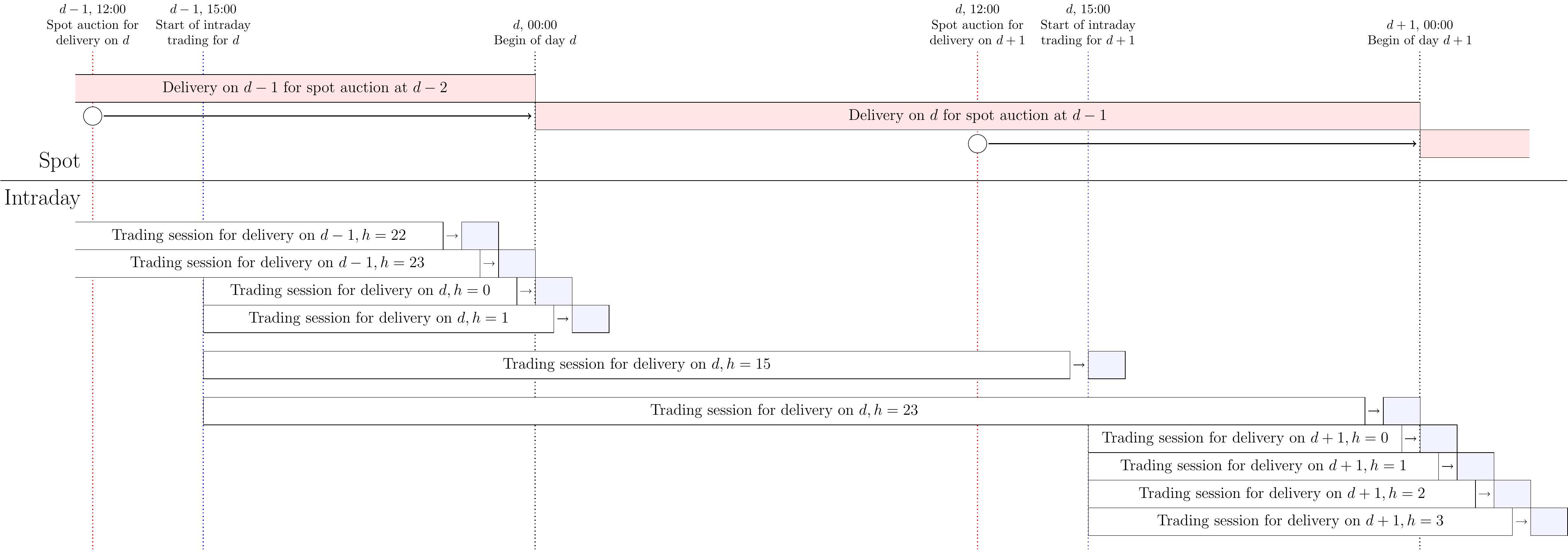}
    \caption{Schedule of the day-ahead spot and the continuous intraday market in comparison. White circles indicate the order book closing of the day-ahead auctions. White boxes indicate trading sessions for different delivery periods. Filled boxes indicate delivery periods. Own figure based on the schedules given in \cite{epex2021trading}. Note that we have omitted trading windows during days to save space.}
    \label{fig:market_structure_shortterm_parallel}
\end{figure}

The intraday market is structured as continuous forward market. For all hourly delivery periods with delivery on $d$, trading starts at $d-1$, 15:00 hours and ends 5 minutes before the actual start of delivery. Within the Single Intraday Coupling (SIDC), the order books of all major continuous intraday markets across Europe are coupled, as long as there is sufficient cross-border transmission capacity. The coupling proceeds in two waves: first, at $d-1$, 18:00 hours, the order books of Germany, Denmark, Sweden, Poland and Norway and Netherlands\footnote{Norway and Netherlands are coupled through the NorNed high-voltage submarine cable. Within the central European Core region, the Netherlands is coupled to the geographic neighbours at 22:00 hours.} are coupled. At $d-1$, 22:00, France, Netherlands, Belgium, Austria, the Czech Republic, Hungary, Romania follow \cite{nemo2021single}. All shared order books close 60 minutes before the start of physical delivery and trading resumes with Germany wide delivery. 30 minutes before the start of physical delivery, the Germany wide delivery and trading resumes on a TSO/grid zone-level. Finally, 5 minutes before delivery, the grid-zone trading closes as well. Note that all hourly (and also half-hourly and quarter-hourly) delivery periods for a delivery day $d$ are traded in parallel, as it shown in Figure \ref{fig:market_structure_shortterm_parallel}. A more detailed overview on intraday electricity markets can be found in \cite{shinde2019literature} and \cite{viehmann2017state}.

\section{Data}\label{sec:data}

The following chapter gives a brief introduction of the data used in this paper and the required pre-processing. We use intraday transaction data from EPEX, the anonymous day-ahead spot auction bid curves from EPEX, and wind, solar and demand forecasts from SMARD respectively ENTSO-E. We also provide summary statistics. As a rule of thumb, superscript indices denote delivery periods while subscript indices denote trading time. We hope this makes the forward market structure of the short-term electricity markets more clear.

\begin{figure}
    \centering
    \includegraphics[width=0.75\textwidth]{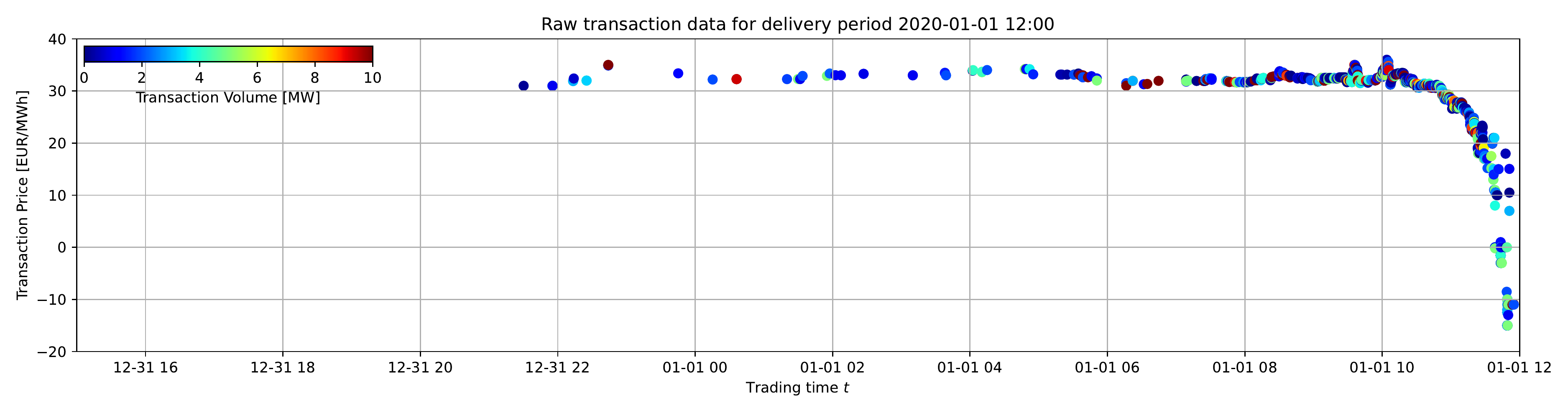}
    \includegraphics[width=0.75\textwidth]{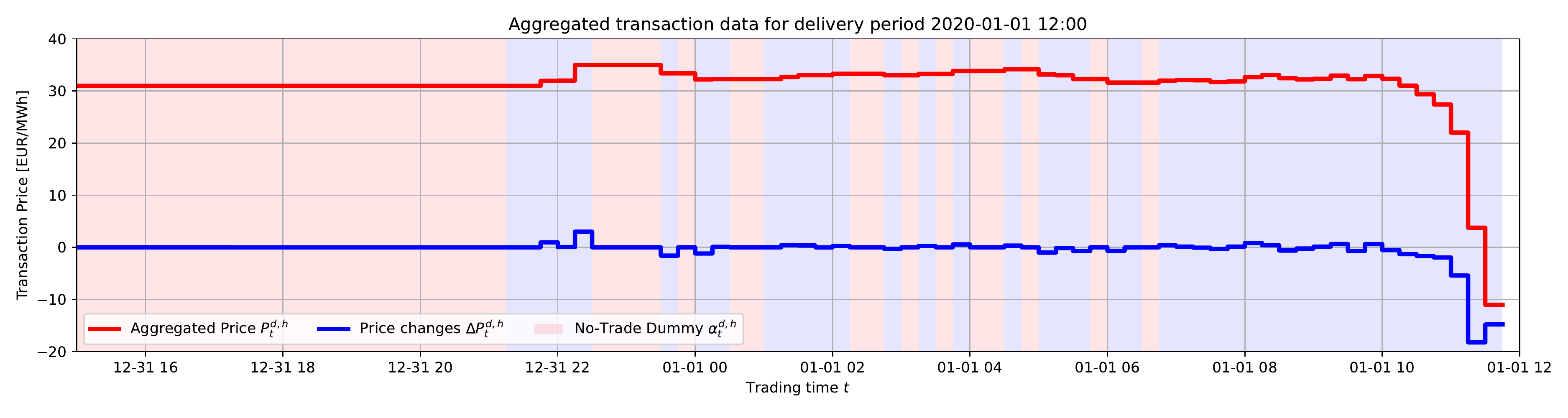}
    \caption{Data aggregation process. The top panel shows the raw transaction data. The lower panel shows the volume weighted 15-minute price path $P_t^{d,h}$ (red line) and the price differences $\Delta P_t^{d,h}$ (blue line). The background shading indicates the Boolean variable $\alpha_t^{d,h}$, where red denotes no trades. The trading periods until the first trade are filled with $P_\text{Spot}^{d,h}$. Own figure.}
    \label{fig:price_data_preparation}
\end{figure}

Intraday transactions are individual trades conducted on the continuous intraday market. We use only trades conducted with either the buy- or sell leg in one of the 4 German grid zones. As trading happens continuously in the intraday market, transactions are irregular spaced in time and need to be aggregated. In line with \cite{hirsch2022simulation, narajewski2020ensemble, narajewski2020econometric, serafin2022trading}, we aggregate all trades for delivery period $d, h$ on a 15-minute equidistant grid along the trading time (denoted with $t$), where $t=0$ denotes the first 15 minutes after trading start. Let $P_t^{d,h}$ denote the volume-weighted average price of all trades with delivery period $d, h$ belonging to bucket $t$ and $\alpha_t^{d,h}$ denote a Boolean indicator whether there was at least one trade. The spot price for each delivery period is denote as  $P_\text{Spot}^{d,h}$. We drop all trades in the local trading phase in the last 30 minutes to the start of physical delivery. The full aggregation process can be seen in Figure \ref{fig:price_data_preparation}. Summary statistics for the price differences are given in Table \ref{tab:price_data_sumamry_stats}. We especially note an increasing volatility in year 2022 driven by the Russian invasion in Ukraine and the energy crisis in Europe. Additionally, we note that trading activity in general increases as the share of no-trade event decreases. Figure \ref{fig:price_data_correlation} gives the pairwise correlation between price changes for all $24 \times 24$ intraday delivery periods during the training set. We already note a cluster of high correlation for the night hours and for the afternoon peak hours.

\begin{minipage}{0.4\textwidth}
   \begin{table}[H]
        \centering
        \begin{tabular}{lrrr}
            \toprule
            {} &    2020 &    2021 &     2022 \\
            \midrule
            Count &  445184 &  480948 &   505673 \\
            Mean  &    0.03 &    0.16 &     0.13 \\
            Std   &    4.33 &    6.46 &    13.12 \\
            MAD   &    0.62 &    1.03 &     2.51 \\
            IQR   &    1.24 &    2.06 &     5.02 \\
            Min   & -999.35 & -523.97 & -1600.27 \\
            Q5\%   &   -3.06 &   -5.61 &   -11.70 \\
            Q10\%  &   -1.75 &   -3.06 &    -6.89 \\
            Q25\%  &   -0.62 &   -1.00 &    -2.50 \\
            Q50\%  &       0 &       0 &        0 \\
            Q75\%  &    0.63 &    1.06 &     2.52 \\
            Q90\%  &    1.75 &    3.25 &     6.87 \\
            Q95\%  &    3.06 &    6.12 &    11.71 \\
            Max   & 1007.97 &  735.88 &  2356.06 \\
            \bottomrule
        \end{tabular}
        \caption{Summary Statistics for all ${\Delta P_t^{d,h} \mid \alpha_t^{d,h} = 1}$. MAD denotes the median absolute deviation and IQR denotes the interquartile range.}
        \label{tab:price_data_sumamry_stats}
    \end{table} 
\end{minipage}
\begin{minipage}{0.6\textwidth}
    \begin{figure}[H]
        \centering
        \includegraphics[width=\linewidth]{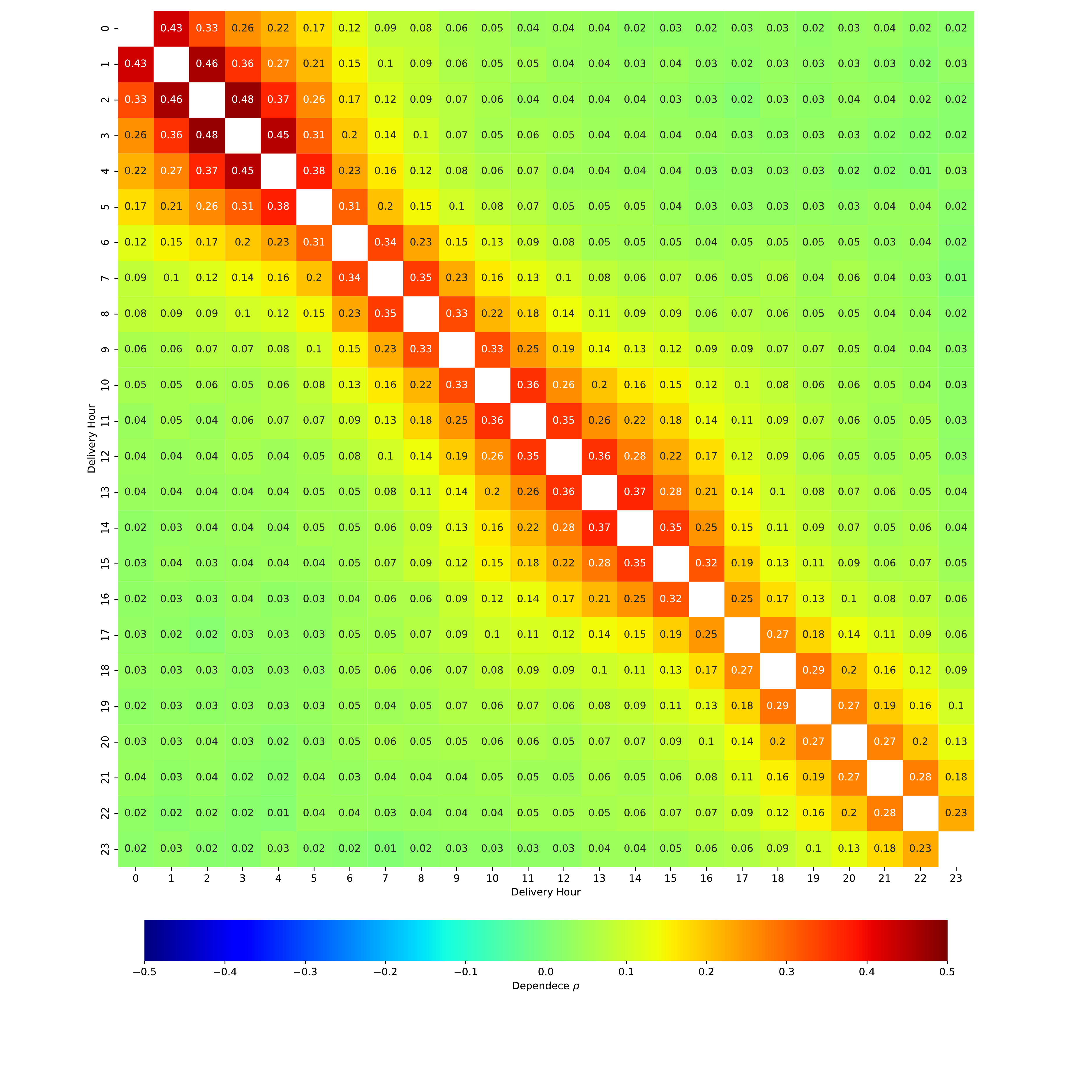}
        \caption{Dependence matrix for the 24-dimensional intraday price change vector $\Delta P_t^{d,h}$. We calculate pairwise correlation to account for the different trading window lengths.}
        \label{fig:price_data_correlation}
    \end{figure}
\end{minipage}

\newcommand{\wion}{\ensuremath{\text{WindOn}^{d,h}}}
\newcommand{\wioff}{\ensuremath{\text{WindOff}^{d,h}}}
\newcommand{\demand}{\ensuremath{\text{Load}^{d,h}}}
\newcommand{\solar}{\ensuremath{\text{Solar}^{d,h}}}

We use wind on- and offshore, solar and demand forecasts from ENTSO-E. The data is aggregated to hourly resolution using a simple arithmetic average. Forecasts are generated by the transmission system operator for each delivery period $d, h$ and available at the day-ahead stage, i.e. latest at $d-1$, 12:00 o'clock. We denote the forecasts as $\wion, \wioff, \solar$ and $\demand$. For all data, we adjust the daylight saving times by (back-) filling the missing hour in spring and averaging the double hour in autumn as it is standard in the electricity price forecasting literature.

\begin{figure}
    \centering
    \includegraphics[width=\textwidth]{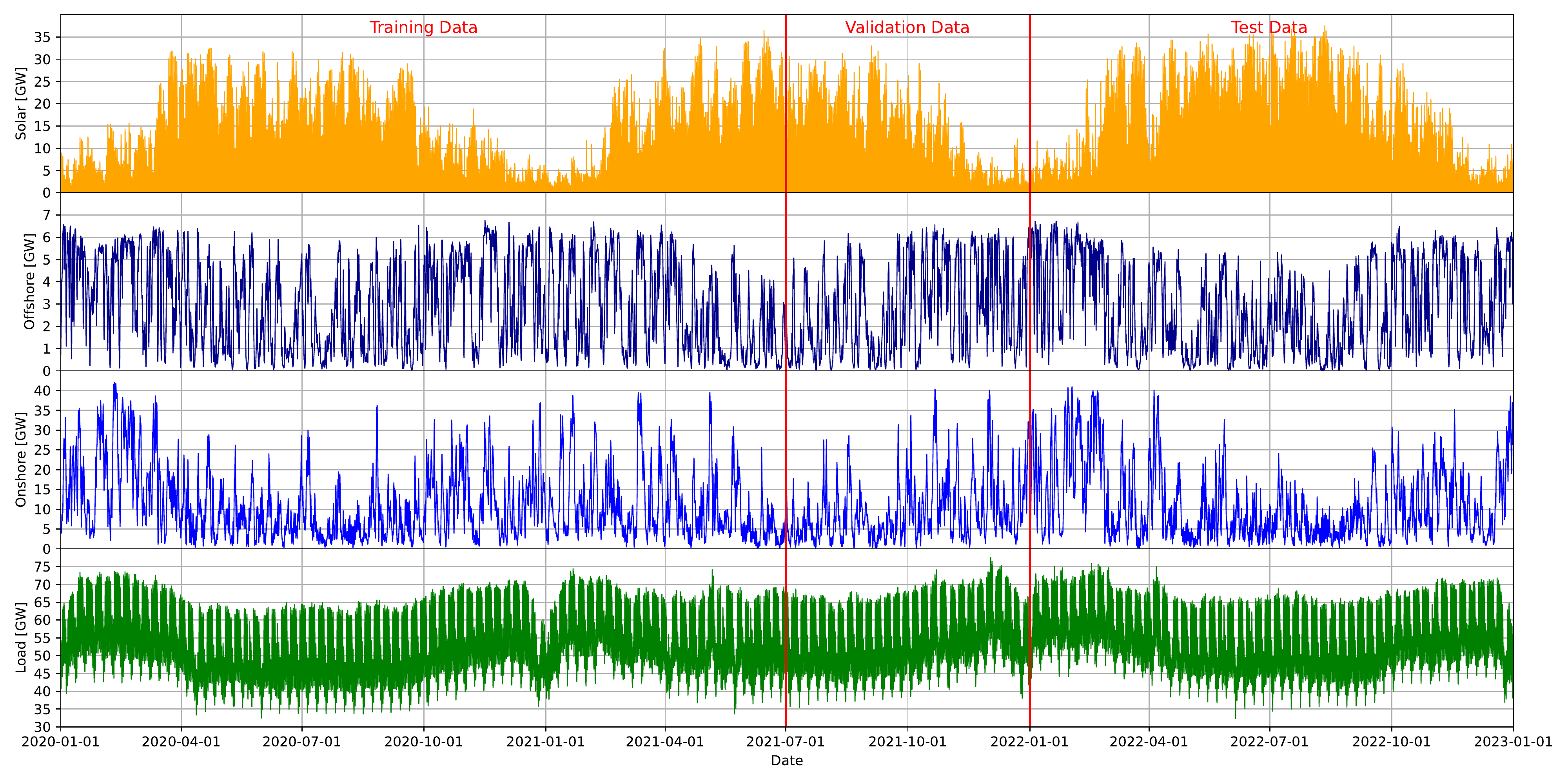}
    \caption{Solar, wind on- and offshore and load day-ahead forecasts from ENTSO-E/SMARD. The initial training data set consists of the first 1.5 years. The subsequent half year is used as validation data set for tuning hyperparameters and model selection. We evaluate our results on the final year of data.}
    \label{fig:data_fundamental_forecasts}
\end{figure}

Additionally, we employ a metric for the merit-order regime. In the classical model, the merit-order is defined as the supply side of the electricity market, sorted by the marginal production costs. The intersection of the supply and demand curve gives the market price. Depending on the slope of the merit-order, changes in the supply or demand have different impacts on the market price. \cite{kremer2021econometric} and \cite{hirsch2022simulation} have shown that the different merit-order regimes explain the size and volatility of price changes in the intraday markets. However, modelling the merit-order in short-term power markets is not straight forward and several approaches have proposed. We follow the approach of \cite{hirsch2022simulation} by using anonymous bid and offer curves from the day-ahead market to model the intraday merit order. The curves are published by EPEX Spot around 14:30 and are therefore available at the time of forecasting. 

\begin{figure}
    \centering
    \includegraphics[width=0.33\textwidth]{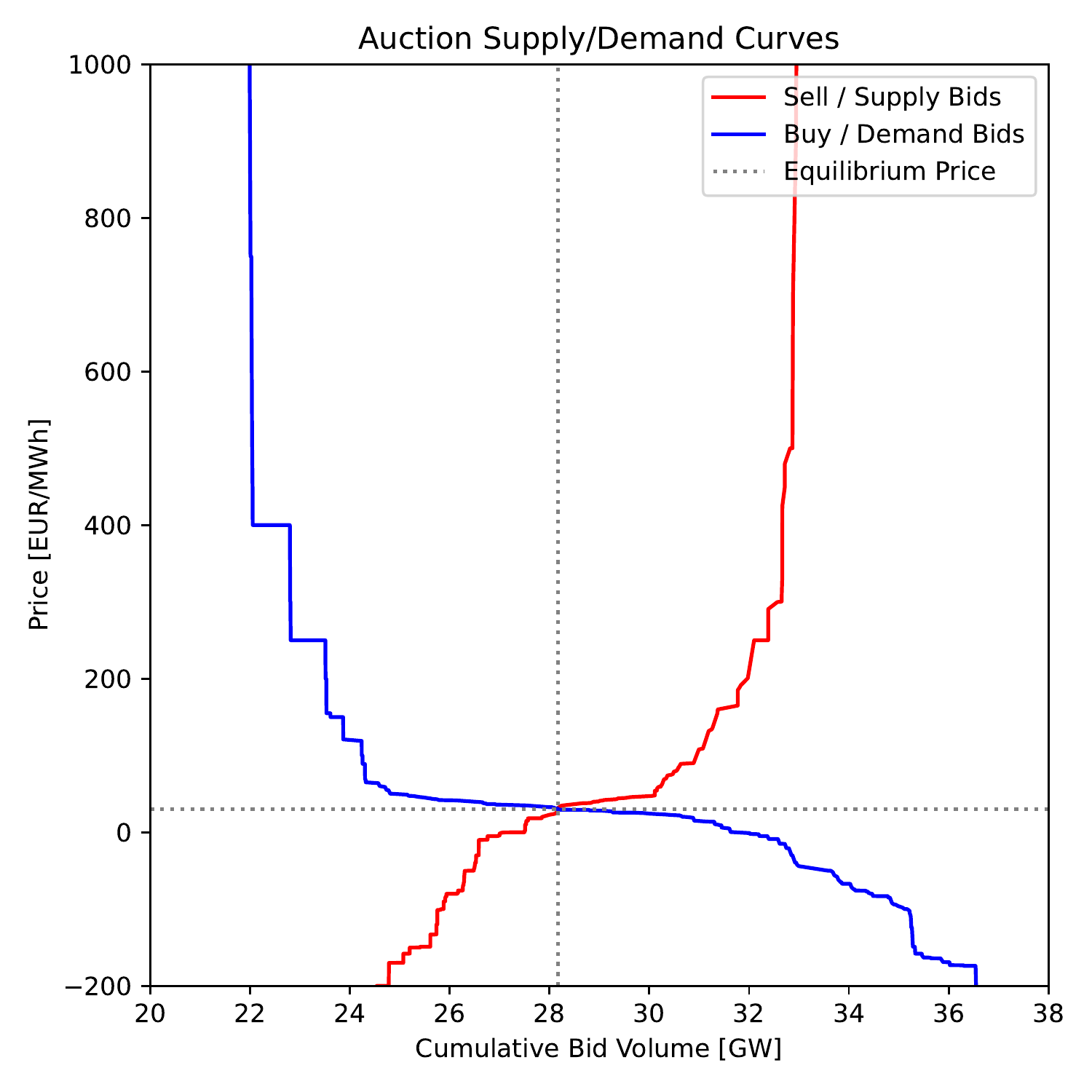}%
    \includegraphics[width=0.33\textwidth]{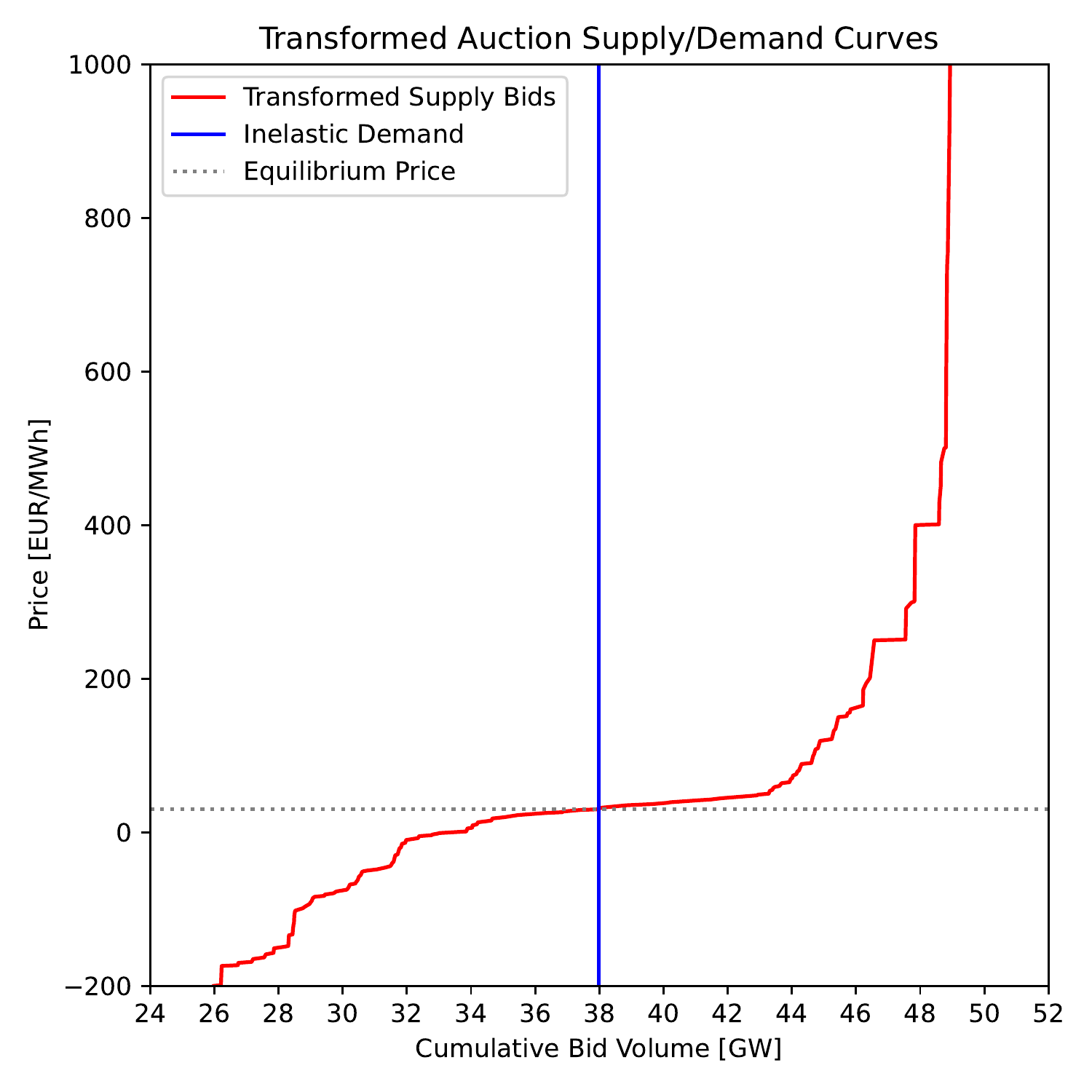}%
    \includegraphics[width=0.33\textwidth]{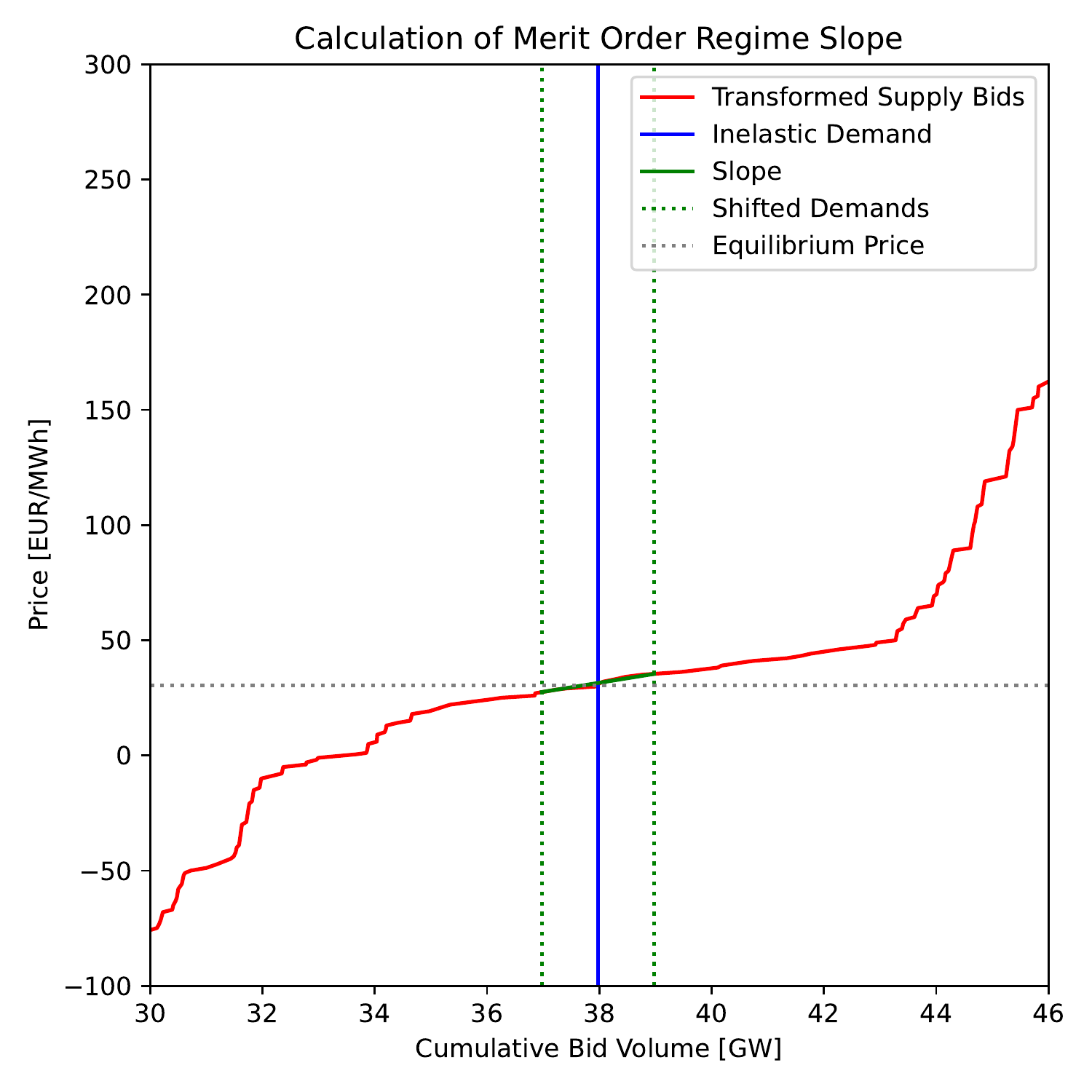}%
    \caption{Calculation of the merit-order regime coefficient $\text{MO}^{d,h}$. The first panel shows the anonymous supply and demand bids for delivery day 2020-01-01 13:00. The second panel shows the transformed auctions curves. The third panel shows the calculation of the slope coefficient (green) around the equilibrium price. Note that we zoomed $x$ and $y$-axes for the third panel.}
    \label{fig:merit_order_slope}
\end{figure}

The overview of the strategy for the calculation of the merit-order regime coefficient $\text{MO}^{d,h}$ is given in Figure \ref{fig:merit_order_slope}. We take the anonymous supply and demand curves from the auction and transform these into an elastic supply curve and an inelastic demand curve using the same transformation as in \cite{hirsch2022simulation, kulakov2021impact}. This transformation is based on the idea that buying 50 MW up to a price of 100 EUR/MWh is the same as buying 50 MW at any price and placing a sell order at 100.01 EUR/MWh. Intuitively, we can therefore create a price independent buy curve and move the price-dependent bids to the sell side. The resulting demand and supply curves are depicted in the middle panel of Figure \ref{fig:merit_order_slope}. Note that the equilibrium price at the intersection between supply and demand curve is unchanged. Lastly, we calculate the slope around the equilibrium price as finite difference quotient. For the exact calculation, we refer the reader to \cite{hirsch2022simulation}.

\section{Models} \label{sec:models}

Our approach follows the widely used inference for margins approach for copula models. We first use a univariate model to estimate the time-varying conditional marginal distribution, apply the probability integral transform and estimate the copula distribution respectively the time-varying dependence parameter in the second step. The following two subsections \ref{sec:models_marginal} and \ref{sec:models_dependence} describe our modelling in more detail. Section \ref{sec:models_simulation} describes the general simulation set-up. Subsection \ref{sec:models_benchmark} describes the different (nested) models for our forecasting study and our benchmark models.

Let us remark that strictly, Sklar's theorem \cite{sklar1973random} is only valid for \emph{continuous} marginal distributions, as $\mathcal{F}_X(X) \sim \mathcal{U}_{[0, 1]}$ is not true for discrete distributions $\mathcal{F}$. This leads to the issue that the copula distribution is not necessarily identifiable with discrete or mixed discrete-continuous marginals. A practical approach to alleviate this issue is to fill the gaps induced through the discreteness of the marginal distribution by some uniform 'jitters', which results in the so-called checkerboard copula, a strategy we employ in this paper as well \cite{geenens2020copula, faugeras2017inference}.

\subsection{Marginal Model for Electricity Prices} \label{sec:models_marginal}

\newcommand{\mixprob}{\ensuremath{\pi_t^{d,h}}}
\newcommand{\sidc}[1]{\ensuremath{\text{SIDC}_\text{#1}(d,h,t)}}

We model the distribution of the price changes $\Delta P_t^{d,h}$ as a mixture distribution $\mathcal{D}^{d,h}_t$ to account for the zero-inflation in price changes. 
\begin{equation}
\Delta P^{d,h}_t \sim \mathcal{D}^{d,h}_t
\end{equation}
 where $\mathcal{D}^{d,h}_t$ is a mixture distribution from a continuous distribution and the Dirac distribution with an atom at 0, denoted as $\delta_0$.
\begin{equation}
\mathcal{D} = (1 - \alpha_t^{d,h}) \delta_0 + \alpha_t^{d,h} \cdot \mathcal{F}^{d,h}_t
\end{equation}
where $\alpha_t^{d,h} = 1$ indicates a trade event and is modelled as an binomial variable $\alpha^{d,h}_t \sim \mathcal{B}^{d,h}_t(\pi_t^{d,h})$. 

A similar approach is taken by \cite{narajewski2020ensemble, hirsch2022simulation}, who estimate the mixture distribution in a two-stage procedure. Owing to stylized facts on intraday electricity prices, \cite{narajewski2020ensemble, hirsch2022simulation} assume $\mathcal{F}$ to follow the (skewed) Student-$t$ or Johnson's $S_U$ distribution. With regards to \cite{hirsch2022simulation} remarks on the issues with estimation stability using the skew $t$-distribution, we generally use Johnson's $S_U$ distribution in this work. We use a two-step estimation procedure for the estimation. First, we estimate the conditional mixing probability $\pi_t^{d,h}$ using regularized logistic regression. Subsequently, we estimate the conditional distribution parameters on all non-zero price changes $\Delta P_t^{d,h} \mid \alpha_t^{d,h} = 1$ using maximum likelihood. The remainder of this section describes our feature engineering, the exact specifications for our models for the conditional mixing probabilities and the conditional distribution parameters as well as our hyperparameter tuning scheme.\footnote{We have also experimented with estimating the zero-inflation in a single estimation step. However, there are two kinds of zero-inflation present in the intraday trade data: We have periods without trades and periods where trading happens, but the trading does not lead to a change in the price. A joint estimation does not differentiate between both effects and provided inferior results in initial testing. Additionally, the estimation of discrete-continuous mixture distributions is not straight-forward using maximum-likelihood as the probability mass and likelihood functions live on different scales.}

Our predictive variables can be grouped into five groups: 
\begin{enumerate}
    \item Fundamental forecasts. We use the day-ahead forecasts for the solar, wind on- and offshore production, the day-ahead load forecast and a measure for the slope of the merit-order to account for different market regimes. 
    \item Time derived dummies. We use dummies for the hour-of-the-day, denoted as HOUR($d,h,t$) and day-of-the-week, denoted as DOW($d, h, t$).
    \item Market structure dummies. We use dummies to distinguish the periods at which the SIDC pan-European order books open and close. The 1st wave opens at 18:00 hours, the 2nd wave opens at 22:00 hours and the order books close 1 hour before the physical delivery. We use an additional dummy to mark the phase where the pan-European order books are open.
    \item Trading time splines. We use ReLU splines to model non-linear effects of the trading time $t$ and the time to delivery $T-t$.
    \item Trading variables. We use first three lagged prices, absolute lagged prices and the first three lagged values of $\alpha_t$ to account for the auto-regressive nature in the distribution parameters. We also use a spline for the level of the spot price, denoted as ReLU($P_\text{Spot}^{d,h}$).
\end{enumerate}
ReLU splines are piecewise linear approximations on the domain of an explanatory variable. For an arbitrary explanatory variable $x$ we define a set of thresholds $\tau \in \mathcal{T}_x$ and  we define the ReLU splines as
\begin{equation}
    \text{ReLU($x, \mathcal{T}_x$)} = \sum_{\tau \in \mathcal{T}_x} \beta_{\tau,x} \cdot \min(x, \tau)
\end{equation}
thereby the clipped values on the domain can contribute with individual slope coefficients. This allows an efficient approximation of non-linear functional relationships while preserving linearity in the coefficients \cite{nair2010rectified}. ReLU splines are especially useful for variables with known and bounded domain such as the trading time $t$ in our application, but can in theory be used for any continuous variable. Even though the thresholds $\mathcal{T}_x$ can be chosen arbitrarily, we generally use an equidistant grid.

\begin{figure}
    \centering
    \begin{subfigure}{0.5\textwidth}
        \includegraphics[width=\linewidth]{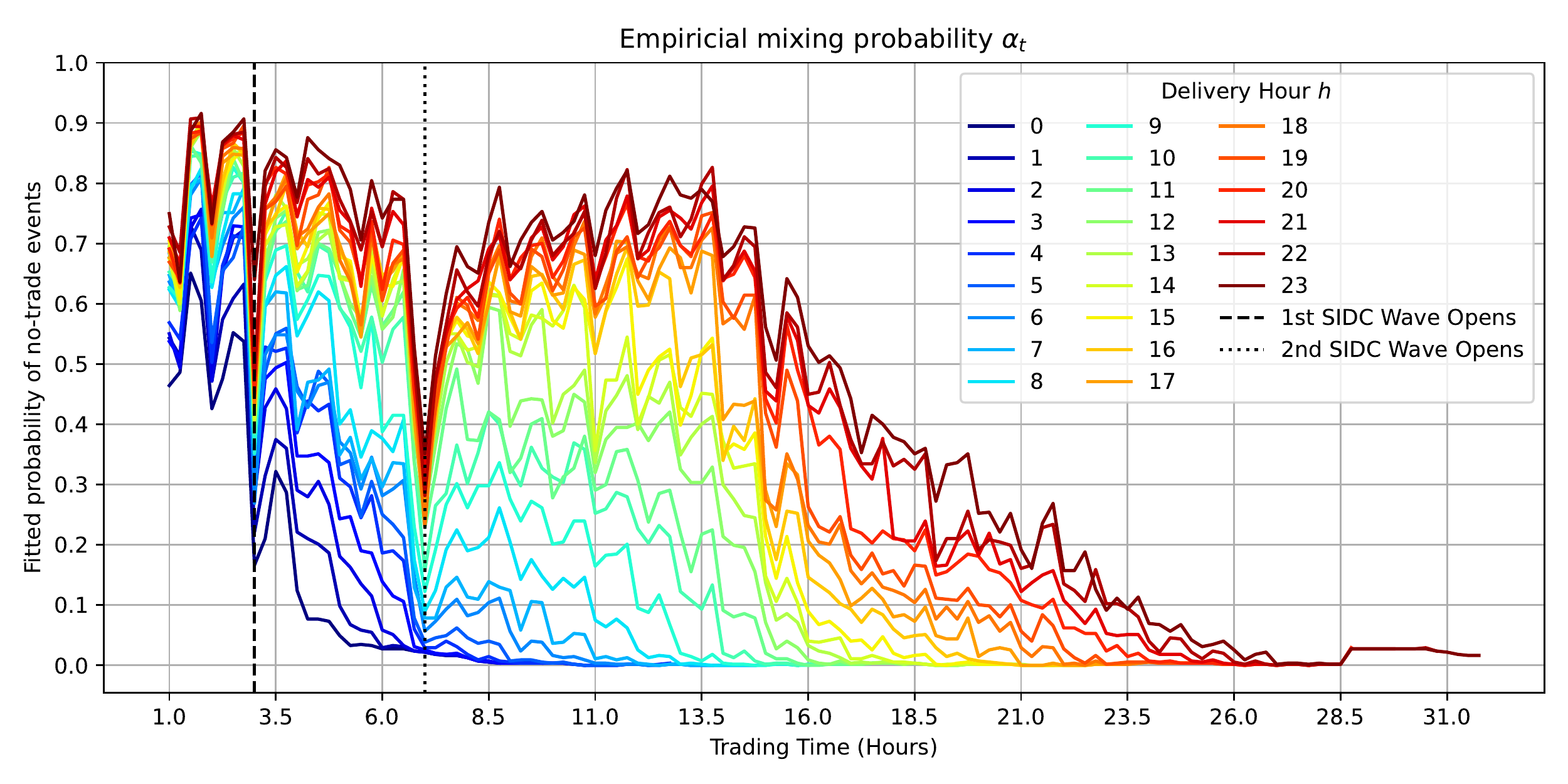}
    \end{subfigure}%
    \begin{subfigure}{0.5\textwidth}
        \includegraphics[width=\linewidth]{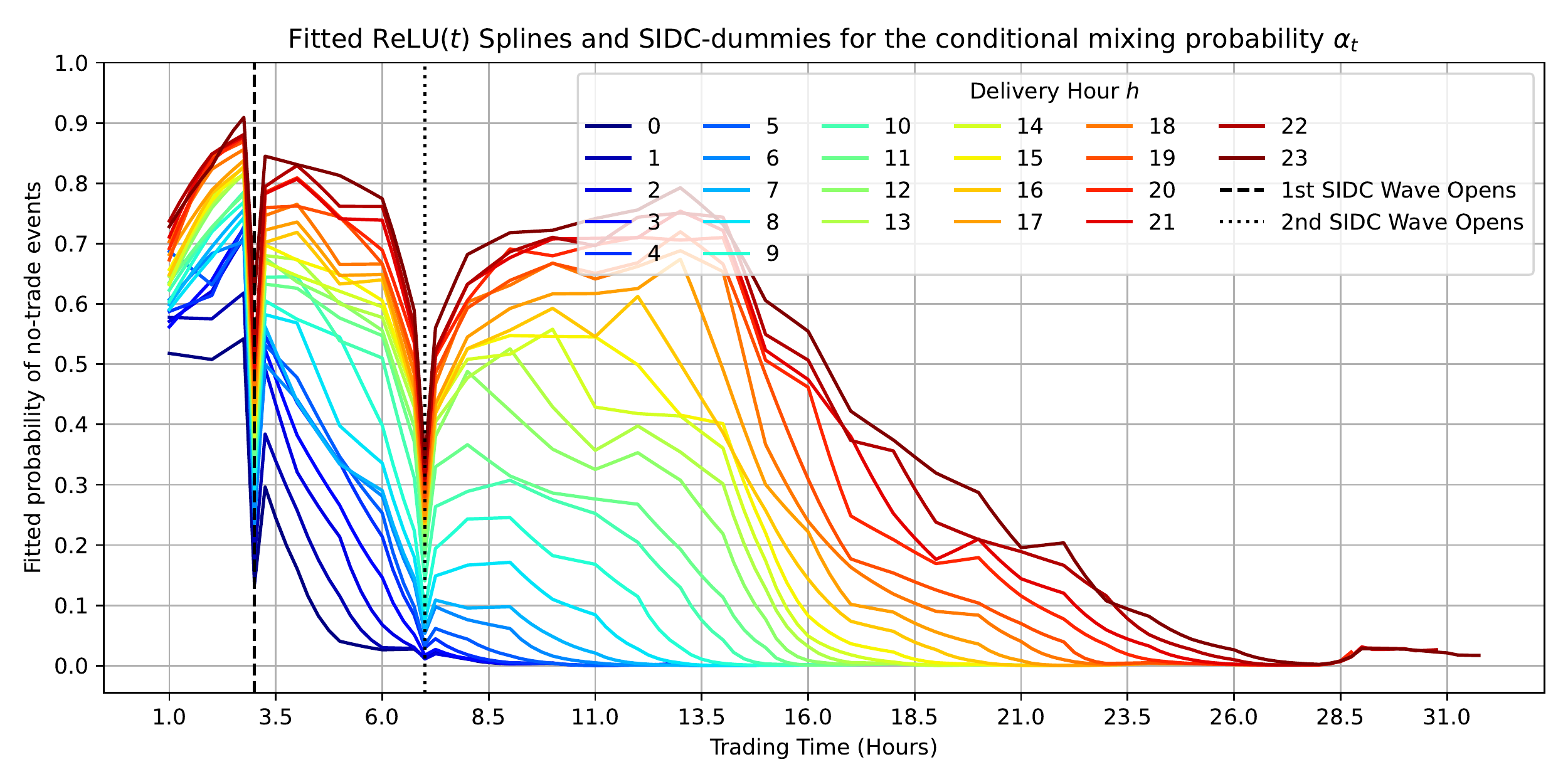}
    \end{subfigure}
    \caption{Empirical mixing probabilities $\alpha_t$ and the fitted conditional mixing probabilities using a ReLU spline for the trading time $t$. Note that in the logistic regression model specified in Equation \ref{eq:logistic_model}, all variables apart from the ReLU spline and the SIDC dummies are constant across $t$, but vary only along $d$ and $s$.}
    \label{fig:fitted_relu_alpha_trading_time}
\end{figure}

The logistic regression model used to estimate the conditional mixing probabilities $\mixprob$ is defined as 
\begin{multline}\label{eq:logistic_model}
        \operatorname{log}\left(\frac{\mixprob}{1-\mixprob}\right) 
         = \beta_0 + \underbrace{\beta_1 \cdot \wion + \beta_2 \cdot \wioff + \beta_3 \cdot \solar + \beta_4 \cdot \demand + \beta_5 \cdot \text{MO}^{d,h}}_{\text{Fundamental Forecasts.}} \\
         + \underbrace{\beta_6 \cdot \sidc{O} + \beta_7 \cdot \sidc{1} + \beta_8 \cdot \sidc{2} + \beta_9 \cdot \sidc{C} +  \beta_{10} \cdot \sidc{L}}_{\text{Dummies for SIDC. See Table \ref{tab:sidc_dummies} for the exact definitions}} \\
         + \underbrace{\sum_{k=1}^6 \beta_{10+k} \cdot \text{DOW}^k(d)}_{\text{Day-of-the-week dummies.}}
         + \underbrace{\sum_{\tau \in \mathcal{T}_t} \sum_h^H \beta_{17 + (\tau,h)} \cdot \text{ReLU}(t, \mathcal{T}_t) \otimes \text{Hour}_h(d, h, s),}_{\text{ReLU splines for trading time per delivery hour.}}
\end{multline} 
where $\beta_0$ denotes the intercept, $\beta_1$ to $\beta_4$ are the coefficients for the fundamental wind, solar and load forecasts, $\beta_5$ to $\beta_9$ are the coefficients for the single intraday coupling dummies $\text{SIDC}_{\{\text{O}, 1, 2, \text{C}, \text{L}\}}(d,h,t)$. The dummies denote whether the cross-border order books are open, the start of the 1st and 2nd wave, the closing of the cross-border order books 1 hour before the gate closure and the last periods of (local) trading. Table \ref{tab:sidc_dummies} gives the specification of the SIDC related dummies. We include day-of-the-week dummies for the delivery day $d$ with the coefficients $\beta_{10}$ to $\beta_{15}$. The last term denotes a ReLU spline for the trading time $t$ on an equidistant grid $\mathcal{T}_t$, with $\otimes$ denoting the Kronecker product. We conduct a grid search on the validation data set to select the step size of the grid $\mathcal{T}_t$ and the regularization parameter based on the validation accuracy. The step size is constant across all $h$. An exemplary fit for the ReLU spline along the trading time $t$ can be seen in Figure \ref{fig:fitted_relu_alpha_trading_time}. The estimation is regularized using the $L_2$ norm, which we choose from an exponential grid on 0.001 to 1000. We also evaluate the Bayesian Information Criterion (BIC) for each combination of step size and $L_2$ regularization and find that the results align.

\begin{table}[]
    \centering
    \begin{tabular}{lll}
        \toprule
        Variable      & Trading Time $t$                  & Interpretation \\
        \midrule
        $\sidc{O}$    & $t \geq 12$ \& $T - t \geq 2$     & All periods in which the cross-border order books are open. \\
        $\sidc{1}$    & $t = 12$                          & First wave of cross-border order book coupling at $d-1$ 18:00 hours. \\
        $\sidc{2}$    & $t = 28$                          & Second wave of cross-border order book coupling at $d-1$ 22:00 hours. \\
        $\sidc{C}$    & $T - t = 2$                       & Closing of the cross-border order books one hour before delivery. \\
        $\sidc{L}$    & $T - t \leq 2$                    & All periods after the closing of cross-border order books. \\
        \bottomrule
    \end{tabular}
    \caption{Specification of the Boolean variables related to the single intraday coupling SIDC. The opening of the cross-border order books happens at 18:00 and 22:00 hours for all delivery periods of the following day. However, the closure of cross-border order books is relative to the delivery time. Hence we need to count backwards in the trading time, taking $T-t$ for variables related to the closure of SIDC.}
    \label{tab:sidc_dummies}
\end{table}

The following paragraph describes the probabilistic model for the marginal distribution $\mathcal{F}$ with time-varying location, scale and shape parameters. Our general framework follows the generalized additive models for location, scale and shape introduced by \cite{rigby2005generalized}. Let $Y = (Y_1, Y_2, ..., Y_n)$ be a vector of $n$ independent observations $Y_i$ and $Y_i$ have the probability (density) function
\begin{equation*}
    f(y_i \mid \mu_i, \sigma_i, \nu_i, \tau_i) = f(y_i \mid \theta_i)
\end{equation*}
where each distribution parameter can be a smooth function of explanatory variables. Denote $\theta_i^k = (\mu_i, \sigma_i, \nu_i, \tau_i) = (\theta^1, \theta^2, \theta^3, \theta^k)$ as the $n \times k$ parameter vector with $k$ location, scale and shape parameters. We have 
\begin{equation}
    Y_i \sim \mathcal{F}(\mu_i, \sigma_i, \nu_i, \tau_i) \Leftrightarrow Y_i \sim \mathcal{F}(\theta_i). 
\end{equation}
Let $g_k(\cdot)$ be a known monotonic link function for each distribution parameter relating $\theta^k$ to explanatory variables through an additive model
\begin{equation}
    g_k(\theta^k) = \eta_k = \boldsymbol{X}_k\boldsymbol{\beta}_k
\end{equation}
where $\boldsymbol{X}_k$ is a $n \times J_k$ known design matrix of $J_k$ exogenous regressors. Additionally, the model can consists of non-linear effects as in classical generalized additive models. Note that each distribution parameter can have an individual design matrix $\boldsymbol{X}_k$.

As noted, we assume Johnson's $S_U$ distribution for the marginal distribution of $\Delta P_t^{d,h}$. The probability density function of Johnson's $S_U$ distribution is defined as follows:
\begin{equation}
    f(y; \mu, \sigma, \nu, \tau) = \frac{1}{\sigma \sqrt{2 \pi}} \frac{\tau}{\sqrt{1 + ((y - \mu) / \sigma))^2}}\operatorname{exp}(-0.5 (\nu + \tau \operatorname{arcsinh}((y - \mu) / \sigma))^2)
\end{equation}
where $-\infty \leq \mu \leq \infty, 0 < \sigma \leq \infty, 0  < \nu \leq \infty, - \infty  \leq \tau \leq \infty$ represent the location, scale, tail and skewness parameters and their respective domains. To ensure all distribution parameters are within their domain, we employ the link functions 
\begin{align}
    g_\mu(x) &= x, \\
    g_\sigma(x) &= \epsilon + \operatorname{log}(1 + \operatorname{exp}(\gamma \cdot x)), \\
    g_\nu(x) &= \epsilon + \operatorname{log}(1 + \operatorname{exp}(\gamma \cdot x)), \\
    g_\tau(x) &= x,
\end{align}
where the link functions for $\mu$ and $\tau$ are the identity and the link functions for $\sigma$ and $\nu$ are known as Softplus link function with constants $\epsilon = 1^{-3}$ and $\gamma = 0.1$ to improve numerical stability \cite{sonnenschein2022probabilistic}.
Formally, we define the model  as follows:
\begin{align}    
    &\begin{aligned}
        g_\mu(\mu) & = \sum_{i=1}^3 \beta_i \cdot \Delta P^{d,h}_{t-i} \\    
    \end{aligned}\\
    &\begin{aligned}
        g_\sigma(\sigma) & = \beta_0 
            + \underbrace{\beta_1 \cdot \wion + \beta_2 \cdot \wioff + \beta_3 \cdot \solar + \beta_4 \cdot \demand + \beta_5 \cdot \text{MO}^{d,h}}_{\text{Fundamental Forecasts.}} \\
            & + \underbrace{\beta_6 \cdot \sidc{O} + \beta_7 \cdot \sidc{1} + \beta_8 \cdot \sidc{2} + \beta_9 \cdot \sidc{C} +  \beta_{10} \cdot \sidc{L}}_{\text{Dummies for SIDC.}} \\
            & + \underbrace{\sum_{k=1}^6 \beta_{10+k} \cdot \text{DOW}^k(d)}_{\text{Day-of-the-week dummies.}}
            + \underbrace{\sum_{h=0}^{23} \beta_{16 + h}\text{Hour}_h(d, h, s)}_{\text{Dummies per delivery hour.}} \\
            & + \underbrace{\sum_{\tau \in \mathcal{T}_t^{10}} \beta_{39 + (\tau,t)} \cdot \text{ReLU}(t, \mathcal{T}_t^{10})}_{\text{Trading time $t$ ReLU spline.}}
            + \underbrace{\sum_{\tau \in \mathcal{T}_{T-t}^{10}} \beta_{52 + (\tau,T-t)} \cdot \text{ReLU}(T-t, \mathcal{T}_{T-t}^{10})}_{\text{Time to delivery $T-t$ ReLU spline.}} \\
            & + \underbrace{\sum_{i=1}^3 \beta_{65+i} \cdot \mid \Delta P_{t-i}^{d,h} \mid}_{\text{Lagged absolute $\Delta P_{t-i}^{d,h}$.}} 
            + \underbrace{\sum_{i=1}^3 \beta_{68+i} \cdot \alpha_{t-i}^{d,h}}_{\text{Lagged $\alpha_{t}^{d,h}$.}} 
            + \underbrace{\sum_{\tau \in \mathcal{T}_\text{Spot}^{50}} \beta_{71 + (\tau,\text{Spot})} \cdot \text{ReLU}(P_\text{Spot}, \mathcal{T}_\text{Spot}^{50})}_{\text{Spot price ReLU spline.}} \\
    \end{aligned}\\
    &\begin{aligned}
        g_\nu(\nu) & =
            \underbrace{\beta_0 \cdot \wion + \beta_1 \cdot \wioff + \beta_2 \cdot \solar + \beta_3 \cdot \demand + \beta_4 \cdot \text{MO}^{d,h}}_{\text{Fundamental Forecasts.}} \\
            & + \underbrace{\beta_5 \cdot \sidc{O} + \beta_6 \cdot \sidc{1} + \beta_7 \cdot \sidc{2} + \beta_8 \cdot \sidc{C} +  \beta_{9} \cdot \sidc{L}}_{\text{Dummies for SIDC.}} \\
            & + \underbrace{\sum_{k=1}^6 \beta_{10+k} \cdot \text{DOW}^k(d)}_{\text{Day-of-the-week dummies.}}
            + \underbrace{\sum_{i=1}^3 \beta_{15+i} \cdot \mid \Delta P_{t-i}^{d,h} \mid}_{\text{Lagged absolute $\Delta P_{t-i}^{d,h}$.}} 
            + \underbrace{\sum_{i=1}^3 \beta_{18+i} \cdot \alpha_{t-i}^{d,h}}_{\text{Lagged $\alpha_{t}^{d,h}$.}} 
            + \underbrace{\sum_{h=0}^{23} \beta_{21 + h}\text{Hour}_h(d, h, s)}_{\text{Dummies per delivery hour.}} \\
    \end{aligned}\\
    &\begin{aligned}
        g_\tau(\tau) & = \underbrace{\sum_{h=0}^{23} \beta_{h}\text{Hour}_h(d, h, s)}_{\text{Dummies per delivery hour.}} 
            + \underbrace{\sum_{k=1}^6 \beta_{23+k} \cdot \text{DOW}^k(d)}_{\text{Day-of-the-week dummies.}} \\
    \end{aligned}
\end{align}
We model the conditional location, scale and shape parameters based on fundamental variables. We guide our selection from the literature \cite{hirsch2022simulation, narajewski2020ensemble, janke2019forecasting, narajewski2020econometric} and regularize the estimation to avoid overfitting.
\begin{itemize}
    \item The location parameter $\mu$ is modelled through three autoregressive lags of $\Delta P_t^{d,h}$.
    \item The scale parameter $\sigma$ is modelled through the the fundamental wind, solar and demand forecasts and the coefficient for the merit-order regime. Additionally, we include day-of-the week and hourly dummies to account for different baseline volatility. We include a ReLU spline for the trading time and the time to delivery to capture the effects of increasing volatility towards gate closure and a ReLU spline for the spot price level to account for the impact of different price regimes. 
    \item The tail parameter $\nu$ is explained using a similar, but slightly reduced set of features as the scale parameter $\sigma$.
    \item The skewness parameter $\tau$ is explained only using day-of-the-week and hourly dummies. This characterization is backed by the recent findings of \cite{hirsch2022simulation}.
\end{itemize}
As the modelling of the distributions higher moments is dependent on the quality of the lower moments' models \cite[on this issue, see. e.g.][]{ziel2022m5}, we refrain from making the models for $\nu$ and $\tau$ as complex as the model for the scale parameter $\sigma$ and decrease complexity accordingly. For the expected value of intraday price changes, many studies have shown indications of weak-form market efficiency \cite{hirsch2022simulation, narajewski2020econometric, narajewski2020ensemble, uniejewski2019understanding, janke2019forecasting, lohndorf2023value}, hence we keep the model simple as possible without comprising the quality of the estimation for $\sigma , \nu$ and $\tau$.

Our estimations are implemented using the \texttt{scikit-learn} \cite{scikit-learn} and the \texttt{tensorflow-probability} packages \cite{dillon2017tensorflow, abadi2016tensorflow}  in \texttt{python}. We automatically tune the hyperparameters of our model using the well-known \texttt{optuna} library, a Bayesian framework specifically tailored towards the optimization of hyperparameters of machine learning models \cite{optuna_2019}. The sampling space for the hyperparameter tuning framework can be found in Table \ref{tab:hyperparameter_search_space}. We initialize the coefficients for the estimation of the conditional location parameter $\mu$ as 0, while the remaining coefficient vectors are initialized uniformly sampled. We tune the $L_1$-regularization for all distribution parameters, the learning rate and introduce a dropout layer during the training process to reduce the risk of overfitting \cite{bengio2012practical, srivastava2014dropout, ziel2022m5}. During model fitting, we reserve 25\% of our training data as validation set to employ early stopping if the training-validation loss does not improve after 25 epochs \cite{ziel2022m5, marcjasz2022distributional}. We run 250 iterations of the \texttt{optuna} algorithm and observe the best trial at iteration 48. Diagnostic plots can be found in the Appendix (see Figure \ref{fig:diagnostics_optuna}).

\begin{table}[h]
    \centering
    \begin{tabular}{lll}
        \toprule
            Parameter                           &   Search space  \\ 
        \midrule
            $L_1$ regularization $\mu$          &   Exponential grid on ($1^{-6}, 10^3$) \\
            $L_1$ regularization $\sigma$       &   Exponential grid on ($1^{-6}, 10^3$) \\
            $L_1$ regularization $\nu$          &   Exponential grid on ($1^{-6}, 10^3$) \\
            $L_1$ regularization $\tau$         &   Exponential grid on ($1^{-6}, 10^3$) \\
            Learning Rate                       &   Exponential grid on ($1^{-5}, 1^{-2}$) \\
            Dropout Rate                        &   Uniform grid on (0, 1) \\ 
        \bottomrule
    \end{tabular}
    \caption{Search spaces for our automated hyperparameter optimization tuning using the \texttt{optuna} framework.}
    \label{tab:hyperparameter_search_space}
\end{table}

\subsection{Dependence}\label{sec:models_dependence}

Remember that we simulate the $T \times H$ price difference vector  $\Delta P^d_t = (\Delta P^{d,1}_t, \Delta P^{d,2}_t, ..., \Delta P^{d,23}_t)$, where the different delivery periods can be correlated. Our strategy follows the general setup of \cite{carmona2021glasso, carmona2022joint}. We estimate the dependence as the covariance after we \emph{gaussianize} our data in the spirit of Sklar's theorem. 

In our copula-based modelling approach, we employ three different structures for the dependence with increasing complexity. The simplest model assumes that all delivery periods are independent and is denoted as \textbf{Mix.Ind}. The next complex model assumes a constant cross-product dependence across the trading window and is denoted as \textbf{Mix.CD}. Lastly, we estimate a time-varying dependence parameter across the trading time $t$. The resulting most complex model is denoted as \textbf{Mix.TD}.

We denote the dependence parameter between two delivery hours $A, B$ as $\rho^{A,B}$. For the constant dependence model, we estimate the pairwise dependence parameter as: 
\begin{equation}
    \rho^{A,B} = \operatorname{cov}(\Delta P_t^{d,A}, \Delta P_t^{d,B}).
\end{equation}
The pair-wise estimation is necessary as not all delivery periods have the same trading length. For our time-varying dependence model \textbf{Mix.TD}, we fit $\rho$ as a latent variable using the beta-regression. 
\begin{equation}
    \rho_{t}^{A,B} = \beta_0 + \sum_{i=1}^{10} \beta_i \cdot \operatorname{ReLU}(t, (0, 12, ..., 128)).
\end{equation}
To estimate the beta-regression, we use the correlation coefficient on the pseudo-Gaussian observations, which we map to the (0, 1) space required by the beta-regression model. We use the BIC to decide on the width of the grid of the spline on $t$ and transform the estimated values back to the covariance using the inverse mapping and the empirical standard deviation of the pseudo-Gaussian observations.

\subsection{Simulation Approach}\label{sec:models_simulation}

\newcommand{\deltapsim}[1]{\ensuremath{\Delta P_{t}^{d,h, [#1]}}}

We simulate $m = 1, ..., M$ paths for the $T \times H$-dimensional intraday price path vector. Denote with $\deltapsim{m}$ the simulated price change at trading time $t$ for delivery period $d, h$. The price path can be seen as the cumulative sum of all price changes and the initial price $P_0^{d,h, [m]}$. In line with previous works \cite[see e.g.][]{nolzen2022market, lohndorf2023value} we initialise the intraday price as the spot price $P_{0}^{d,h,[m]} = P_\text{Spot}^{d,h}$. Formally, we have: 
\begin{equation}
    P_t^{d,h, [m]} = P_\text{Spot}^{d,h} + \sum_{k=1}^t \Delta P_k^{d,h, [m]}
\end{equation}
where it is important to note that the end of trading, $T$, depends on $h$. Therefore, the vector is not square but has the asymmetric trapezoid shape already visible in Figure \ref{fig:trading_in_parallel_trading_sessions} and Figure \ref{fig:market_structure_shortterm_parallel}.

\subsection{Benchmark Models} \label{sec:models_benchmark}

We employ three benchmark models from the literature. First, we employ the well performing \emph{naive} model introduced by \cite{narajewski2020ensemble, hirsch2022simulation}. The model re-uses past price trajectories by sampling. We employ the model in two versions, first assuming independence between different delivery hours (i.e sampling each delivery hour $h = 0, ..., H$ individually) and second by sampling the full $T \times H$ path vector. The third benchmark model is a arithmetic random walk model introduced by \cite{lohndorf2023value} drawing from the empirical price difference distribution. The following paragraphs introduce the benchmark models formally. 

The \textbf{Naive.Ind} and \textbf{Naive.Dep} model is defined as
\begin{equation}
    \Delta P^{d,h, [m]} = \Delta P^{d',h}, 
\end{equation}
where $d'$ is a random day sampled from the training data set $d' \sim \mathcal{U}(\{0, ... , d-1\})$. Note that for the \textbf{Naive.Ind} we sample $d'$ independent for each delivery hour $h = 0, ..., H$ and for the \textbf{Naive.Dep}, we use the same $d'$ for all $h$. \cite{hirsch2022simulation, narajewski2020ensemble} have shown that this type of benchmark models provides very good point and probabilistic forecasting performance in ensemble forecasting settings.

The \textbf{RW.Emp} has been used by \cite{lohndorf2023value} to generate paths for the intraday market in an application study for grid-scale storage optimization. The price process is defined as random walk, where the innovations are drawn from a discrete distribution of the centered empirical price changes. It is defined as:
\begin{equation}
    \Delta P_t^{d,h, [m]} = \Delta P_t^{d',h} - \frac{1}{d-1} \sum_{d=0}^{d-1} (P_{t-1}^{d,h} - P_t^{d,h}),
\end{equation} 
where again $d'$ is a random day sampled from the training data set $d' \sim \mathcal{U}(\{0, ... , d-1\})$ and the second term centers all residuals to mean zero, ensuring the price process to be a martingale. 

For all three benchmark models, the size of the training data $d = 0, ..., d-1$ is a tuning parameter, which we optimize through a grid search.

\section{Forecast Study and Evaluation Metrics} \label{sec:study_design}

We employ the well-known rolling window forecasting study design. Our data comprises of the three years between 2020-01-01 and 2023-01-01. We split our data set in a training, validation and test data set. Our initial training set is 2020-01-01 to 2021-07-01, the validation set contains 2021-07-01 to 2022-01-01 and our test set contains the final year 2022-01-01 to 2023-01-01. The split is also depicted in Figure \ref{fig:data_fundamental_forecasts}. We use the initial training set to develop and estimate our models and the validation data set to calibrate hyperparameters. We use the test set to run a rolling window forecasting study with monthly re-training of all models using the most recent two years of data. This is due to the high computational burden of training and tuning the probabilistic models in \texttt{tensorflow-probability}.

We evaluate our simulations from a point and probabilistic forecasting perspective using strictly proper scoring rules \cite{gneiting2007strictly}. We evaluate the mean and median simulation paths using the well known root mean squared error (RMSE) and median absolute error (MAE). The marginal fit is evaluated using the continuous ranked probability score (CRPS), which we approximate on a dense grid of quantiles using the pinball score (PB). We evaluate the scenario paths using the energy score (ES). The energy score is the multivariate generalization of the continuous ranked probability score, taking into account the correlation structure. We establish significance using the Diebold-Mariano test \cite{diebold2002comparing, diebold2015comparing} for comparing predictive accuracy. The following few paragraphs introduce our metrics formally.

The root mean squared error is defined as:
\begin{equation}
    \text{RMSE}^{d,h} =\sqrt{\frac{1}{T}\sum_{t=1}^T\left(\frac{1}{M} \sum_{m=1}^M P_t^{d,h,[m]} - P_t^{d,h}\right)^2}.
\end{equation}
The RMSE is a strictly proper scoring rule for the expected value. We also report the RMSE averaged additionally across the delivery hours $h$ and delivery days $d$.

Denote the median trajectory over all $M$ simulated trajectories as $\operatorname{med}(P_t^{d,h, [m]})$. The mean absolute error is defined as:
\begin{equation}
    \text{MAE}^{d,h} =\frac{1}{T}\sum_{t=1}^T \mid \operatorname{med}(P_t^{d,h, [m]}) - P_t^{d,h} \mid.
\end{equation}

We evaluate the probabilistic forecasting performance using the continuous ranked probability score (CRPS) and the energy score (ES). Both are  strictly proper scoring rules for the marginal distribution respectively the multidimensional predictive distribution \cite{gneiting2007strictly, ziel2019multivariate} and routinely used in probabilistic energy forecasting \cite{hirsch2022simulation, narajewski2020ensemble, berrisch2023modeling, nowotarski2018recent}. 

The CRPS is approximated using the pinball score on a dense grid of quantiles of our scenarios. Let $Q_{\tau,t}^{d,h}(P_t^{d,h,[m]})$ denote the $\tau$-quantile of our simulation paths $P_t^{d,h,[m]}$. The $\tau$-PB can be defined as:
\begin{equation}
    \text{PB}_{\tau,t}^{d,h} = \begin{cases}
        (1 - \tau) \cdot \left( Q_{\tau,t}^{d,h}(P_t^{d,h,[m]})  - P_t^{d,h} \right)     & \text{for $P_t^{d,h} \leq  Q_{\tau,t}^{d,h}(P_t^{d,h,[m]})$.} \\
        \tau \cdot \left( P_t^{d,h} -  Q_{\tau,t}^{d,h}(P_t^{d,h,[m]}) \right)           & \text{else.}
    \end{cases}
\end{equation}
The CRPS is the average across all quantile levels $\mathcal{T} = \{ 0.01, .., 0.99 \}$ of length 99:
\begin{equation}
    \text{CRPS}^{d,h} = \frac{1}{T}\frac{1}{99} \sum_{t=0}^T \sum_{\tau \in \mathcal{T}} \text{PB}_{\tau,t}^{d,h}
\end{equation}

For the energy score, we implement the $K$-band estimator as given in \cite{ziel2019multivariate}:
\begin{equation}
    \text{ES}_{K}^{d,h} = 
        \frac{1}{M} \sum_{m=1}^M \left\lVert P_t^{d,h,[m]} - P_t^{d,h}\right\rVert_{2} -
        \frac{1}{M \cdot (K-1)} \sum_{m=1}^M \sum_{k=m}^K \left\lVert P_t^{d,h,[m]} - P_t^{d,h,[k+1]}\right\rVert_{2}
\end{equation}
for an integer $1 \leq K \leq M$ and where we set $P_t^{d,h,[M+k]} = P_t^{d,h,[k]}$. $\left\lVert \cdot \right\rVert_{2}$ denotes the $L_2$ or Euclidean norm. We use $K = 10$ as trade-off between computational complexity and estimation accuracy. We evaluate the energy score for the full scenario trajectory and for the last three hours of before the start of physical delivery for each model. The later evaluation acknowledges the importance of the last hours of trading and appeals to practitioners in the field. 

We use the Diebold-Mariano (DM) test to compare the predictive accuracy of the forecasts \cite{nowotarski2018recent, diebold2002comparing, diebold2015comparing}. Intuitively, the DM-test evaluates the null hypothesis ($H_0$) that the difference in means between the loss series of two models is statistically significantly different from zero. Formally, for two models $A$ and $B$, let $L_A$ and $L_B$ the loss series. The loss differential $\Delta_{A,B}$ is defined as
\begin{equation}
    \Delta_{A,B}^i = \left\lVert L_A \right\rVert_{i} - \left\lVert L_A \right\rVert_{i}
\end{equation}
for the $i$-norm. For each model pair, we test two one-sided tests for the null hypothesis (1) $\operatorname{E}\left[ \Delta_{A,B}^i \right] > 0$ and (2) $\operatorname{\mathbb{E}}\left[ \Delta_{A,B}^i \right] < 0$, i.e. (1) the forecasts of model $B$ outperform the forecasts of model $A$ and (2) the forecasts of model $A$ outperform the forecasts of model $B$. These tests are complimentary. Note that the Diebold-Mariano test assumes the loss differential series to be stationary. We test this assumption using the augmented Dickey-Fuller test \cite{dickey1979distribution}.

\section{Results and Discussion} \label{sec:results}

Our parametric approach to modelling the distribution parameters allows us to derive some fundamental insight in the driving factors of the intraday price process. The following section presents first presents some in-sample results from our modelling and subsequently presents the results from our forecasting study. 

\subsection{Fundamental Analysis}

Our parametric modelling set-up allows us to analyse the influence of the driving factors for the location, shape and scale parameters of the price distribution. Our focus here is on the impact of the SIDC on the trading activity and the impact of fundamental variables on the volatility. 

The evolution of the trading activity respectively the share of no-trade events can be seen in Figure \ref{fig:fitted_relu_alpha_trading_time}. We note that in the first hours of trading and that trading activity rises non-linearly towards gate closure. The opening of SIDC induces to spikes in trading activity when the cross-border order books are coupled. At this point, orders in markets with previously different price levels are instantly matched, leading high trading activity for short periods of time. After the SIDC coupling, trading activity increases towards the gate closure. The last trading periods have (almost) no no-trade events. 

\begin{figure}[]
    \centering
    \includegraphics[width=0.5\textwidth]{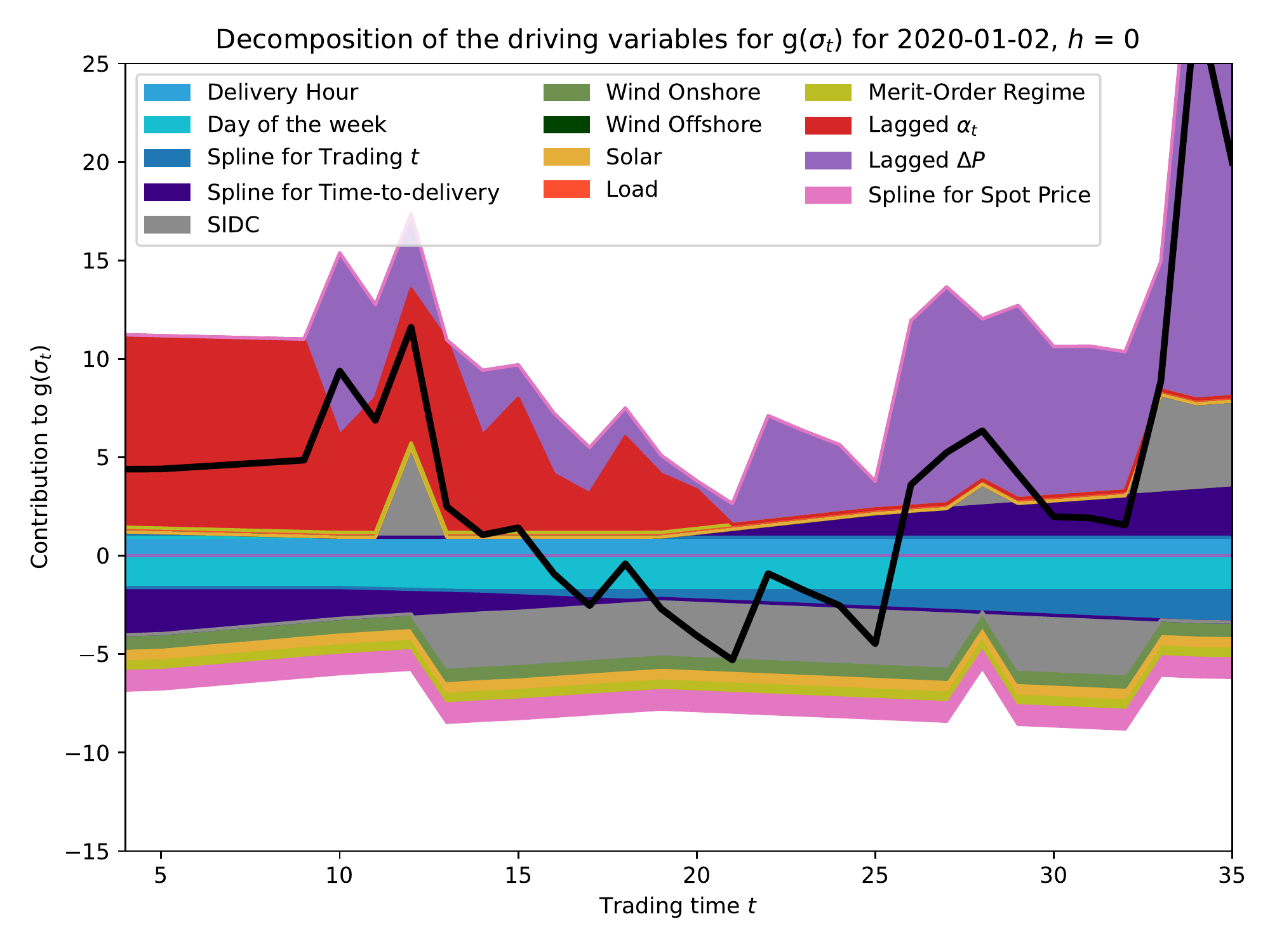}%
    \includegraphics[width=0.5\textwidth]{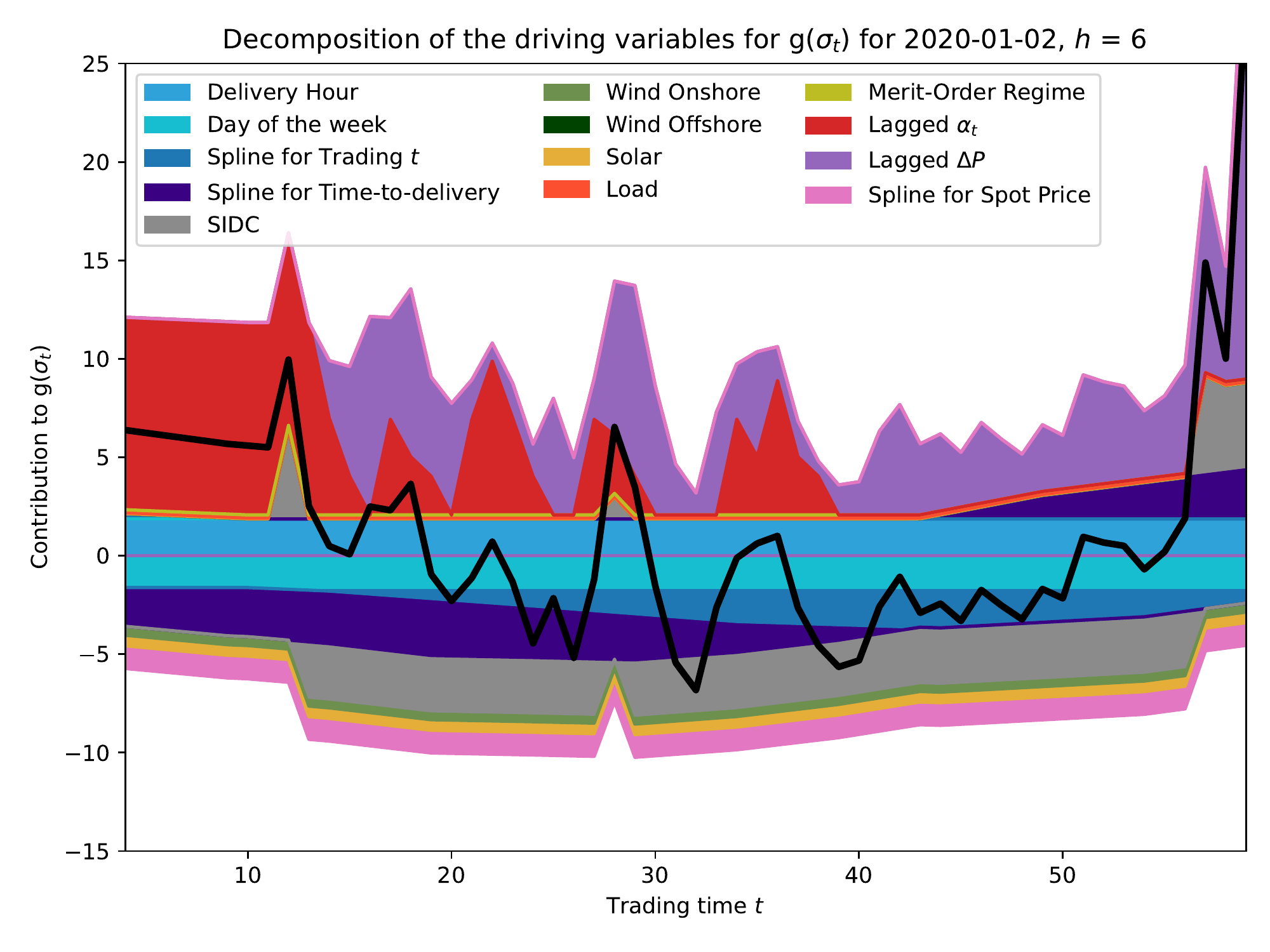}
    \includegraphics[width=0.5\textwidth]{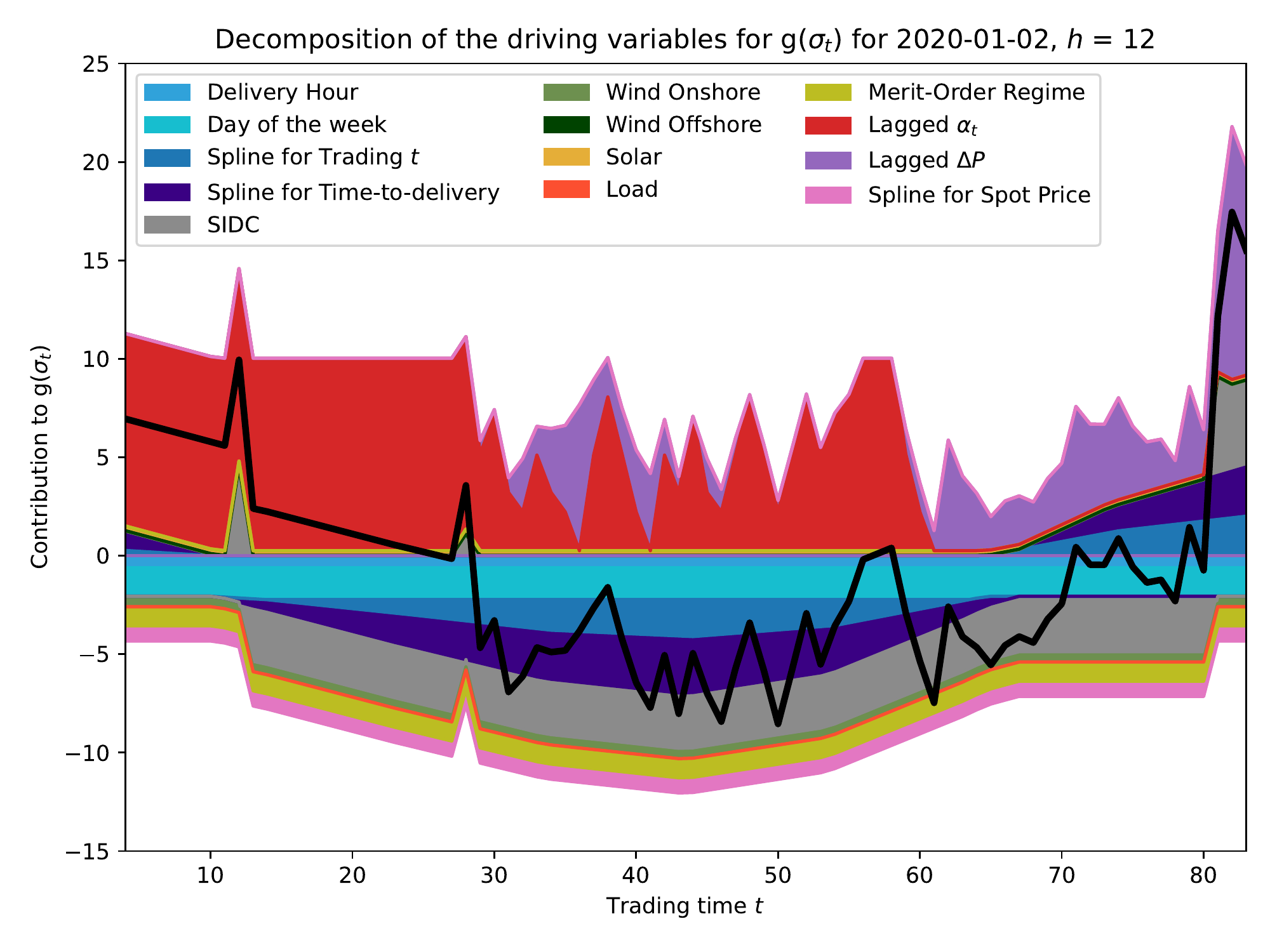}%
    \includegraphics[width=0.5\textwidth]{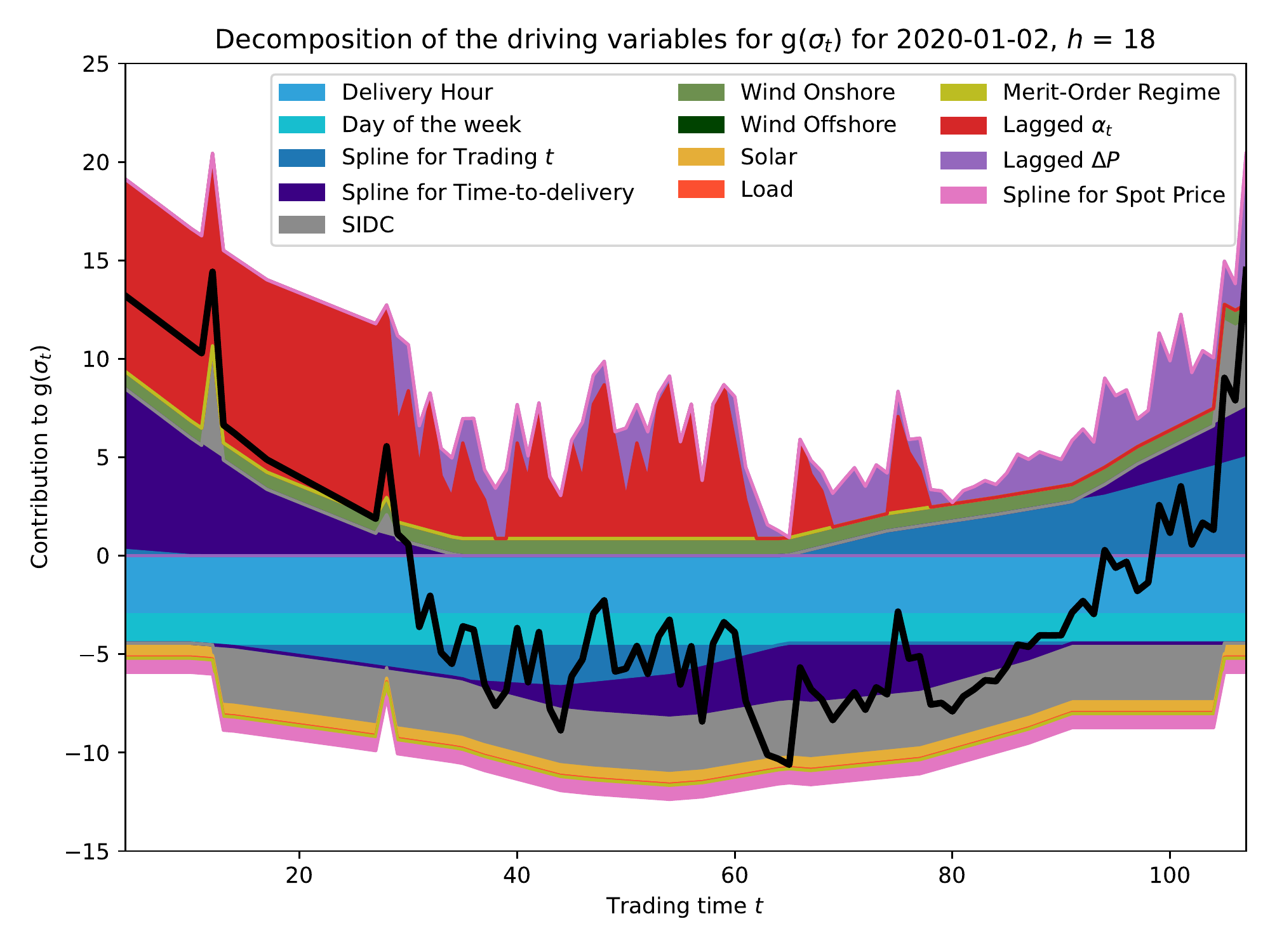}
    \caption{Decomposition of the driving variables for the scale parameter $g_\sigma(\sigma_t^{d, h})$. We show 2020-01-02 for hours 0, 6, 12 and 18. Variables are grouped by color. The impact of fundamental forecasts is constant throughout the trading window as these are only available at the day-ahead stage. The link-function $g_\sigma(\cdot)$ ensures that the final estimated $\sigma^{d,h}_t$ is positive.}
    \label{fig:fundamental_analysis_contribution_scale}
\end{figure}

We show the contribution of individual groups of regressors on the scale parameter in Figure \ref{fig:fundamental_analysis_contribution_scale}. Recall that we model the volatility by four main groups of regressors: fundamental forecasts, time-derived variables, SIDC-related variables, and variables related to the trading activity. Generally, we note a pattern of higher volatility in the beginning of the trading session, followed by decrease and an increase closer to delivery. We note that SIDC has a  distinct impact on the scale parameter: The opening of the cross-border order books at 18:00 and 22:00 for the leads to  clearly visible spikes in the volatility. Subsequently, the phase during which the order-books are coupled is characterized by lower volatility. The closing of the cross-border order books shortly before delivery leads to a spike in volatility. The dampening effect of the open SIDC order books on volatility is likely due to the increased liquidity available to market participants, while the volatility spikes during the opening at 18:00 and 22:00 hours can be explained by matching the order books at different price levels. Overall, the effects contradict the results of \cite{kath2019modeling}, who finds no effects of SIDC, but align with \cite{narajewski2020ensemble, hirsch2022simulation} on the effect of the SIDC closing period. Comparing the different delivery hours $h = 0, 6, 12$ and $18$, we note the different impact of the time to delivery and trading time splines. The time-to-delivery spline kicks in early, but keeps rather constant shortly before the delivery. On the other side, the trading time $t$ spline rises with trading time.

The influence of fundamental variables like wind, solar and demand forecasts is small. One reason for this might be, that these forecasts are generated at the day-ahead stage and are not updated throughout the trading window, as previous works \cite{kremer2021econometric, ziel2017modeling, hirsch2022simulation} have shown that the intraday market is more impacted by forecast changes compared, while the level of forecasts is less important. Additionally, the conclusion that fundamental variables seem not to improve forecasts supports the notion of market efficiency as already indicated by \cite{narajewski2020econometric, narajewski2020ensemble, hirsch2022simulation}. 

\subsection{Forecasting Performance}

\begin{figure}
    \centering
    \begin{minipage}{0.33\textwidth}
        \includegraphics[width=\linewidth]{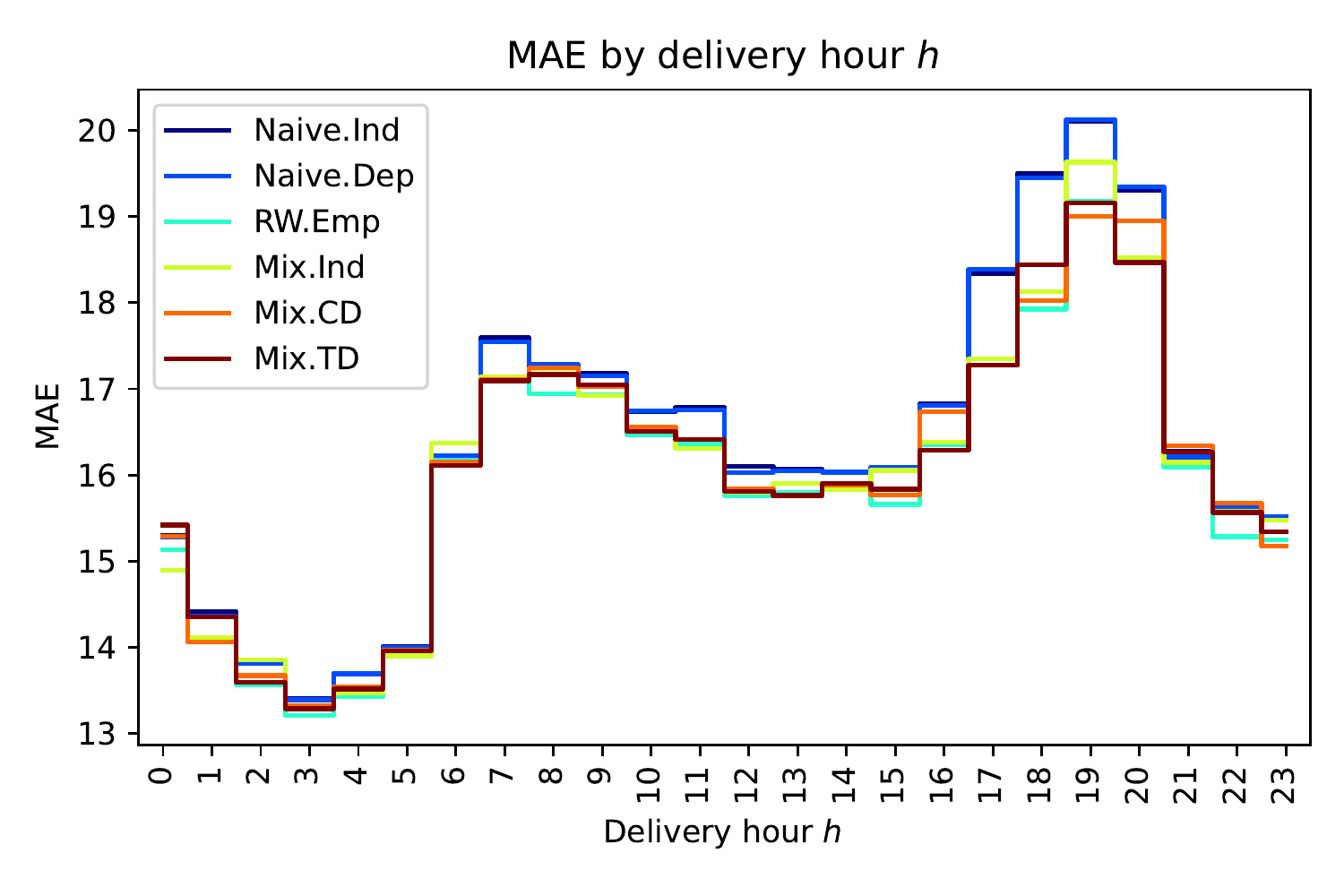}
    \end{minipage}%
    \begin{minipage}{0.33\textwidth}
        \includegraphics[width=\linewidth]{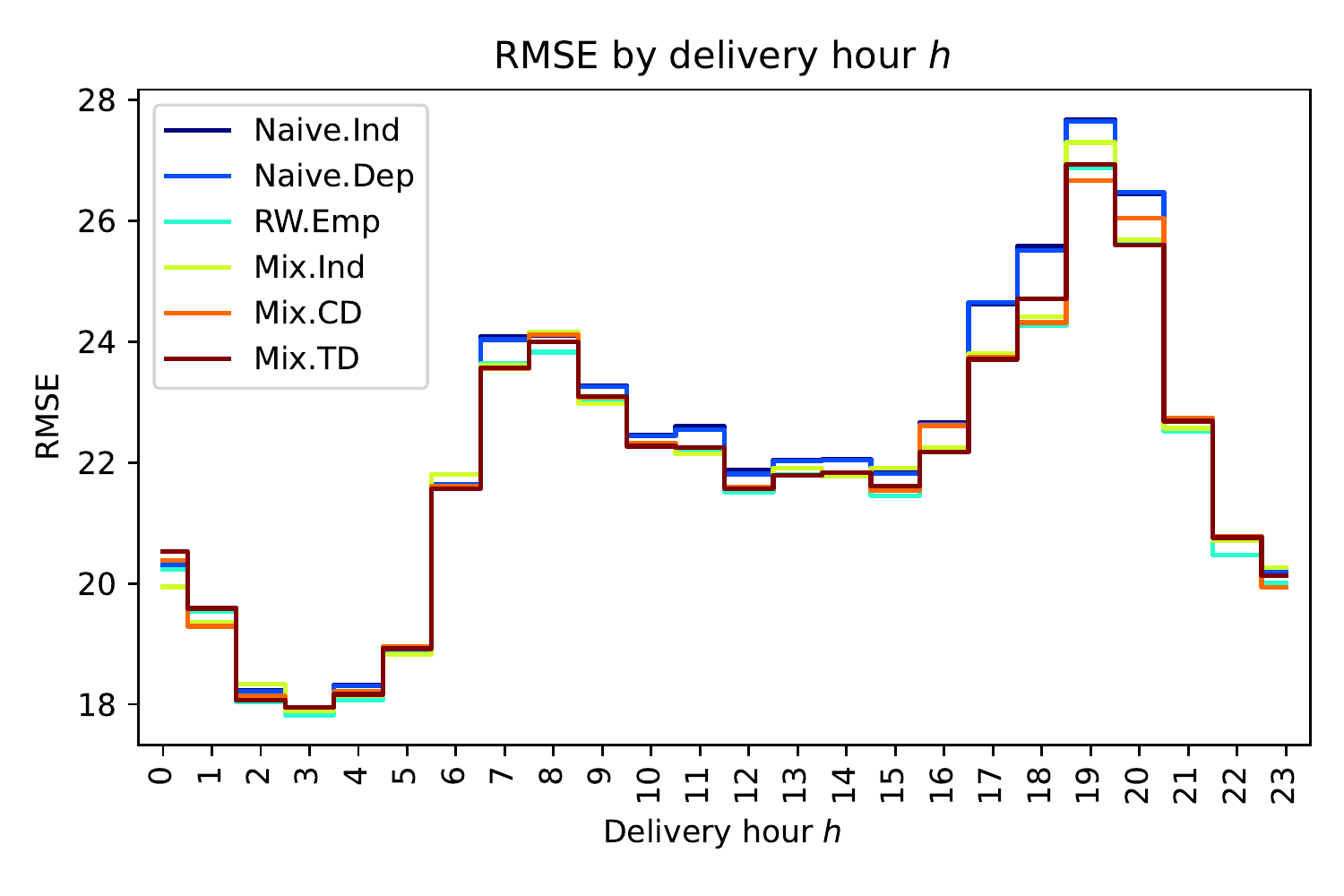}
    \end{minipage}%
    \begin{minipage}{0.33\textwidth}
        \includegraphics[width=\linewidth]{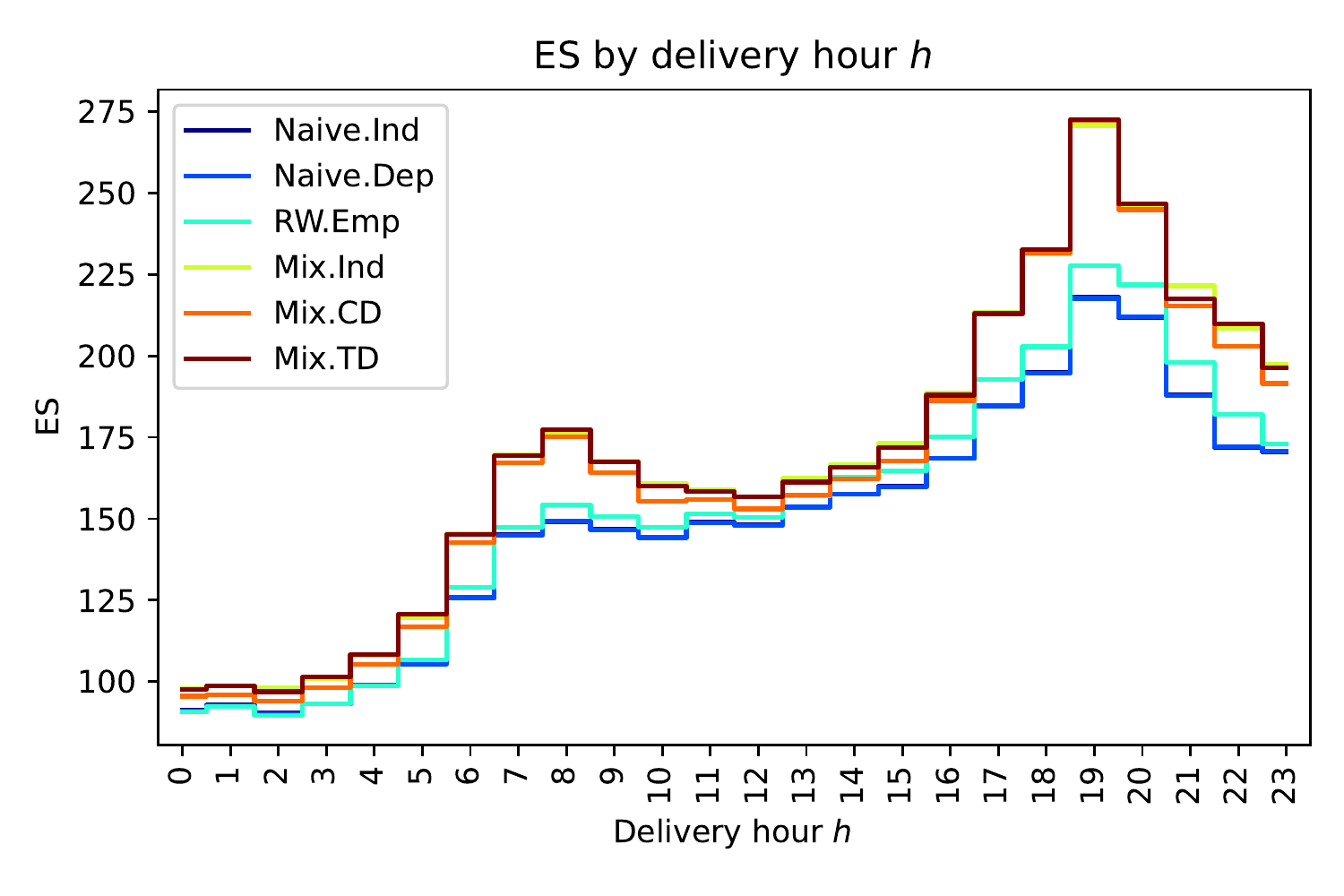}
    \end{minipage}%
    \caption{Error statistics by delivery hour $h$. Note that the scale of the MAE and RMSE is not directly comparable to the ES.}
    \label{fig:errors_by_hour}
\end{figure}

\begin{figure}
    \centering
    \includegraphics[width=0.5\textwidth]{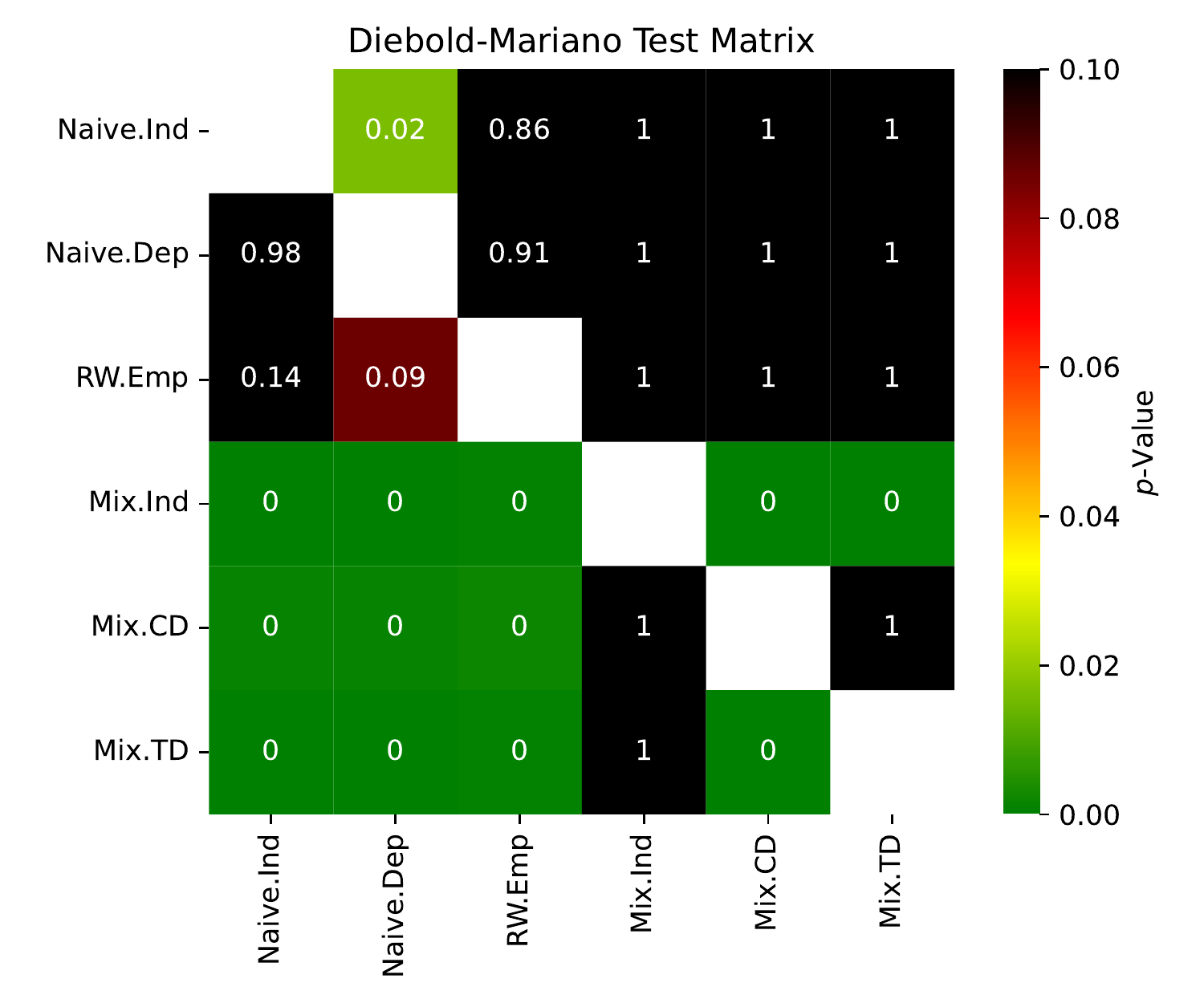}
    \caption{Diebold-Mariano Test Matrix. A $p$-value $< 0.05$ implies that the model on the column has significantly better forecasts than the model on the row.}
    \label{fig:errors_dmtest}
\end{figure}

This section presents the results of the out-of-sample forecasting study. Aggregate error metrics are given in Table \ref{tab:errors_scoring_rules}. Figure \ref{fig:errors_by_hour} presents the error metrics by the delivery hour $h$. Figure \ref{fig:errors_dmtest} gives the results of our pairwise Diebold-Mariano tests. 

\begin{table}[]
    \centering
        \begin{tabular}{lrrrrr}
        \toprule
            & MAE & RMSE & CRPS & ES & $\text{ES}_\text{3H}$ \\ 
        \midrule
            Naive.Ind & {\cellcolor[HTML]{F0F921}} \color[HTML]{000000} 16.340 & {\cellcolor[HTML]{F0F921}} \color[HTML]{000000} 26.649 & {\cellcolor[HTML]{F0F921}} \color[HTML]{000000} 7.470 & {\cellcolor[HTML]{2A0593}} \color[HTML]{F1F1F1} 812.705 & {\cellcolor[HTML]{370499}} \color[HTML]{F1F1F1} 191.423 \\
            Naive.Dep & {\cellcolor[HTML]{F3EE27}} \color[HTML]{000000} 16.330 & {\cellcolor[HTML]{FCD025}} \color[HTML]{000000} 26.626 & {\cellcolor[HTML]{AC2694}} \color[HTML]{F1F1F1} 6.782 & {\cellcolor[HTML]{0D0887}} \color[HTML]{F1F1F1} \bfseries 801.908 & {\cellcolor[HTML]{0D0887}} \color[HTML]{F1F1F1} \bfseries 185.909 \\
            RW.Emp & {\cellcolor[HTML]{0D0887}} \color[HTML]{F1F1F1} \bfseries 15.942 & {\cellcolor[HTML]{0D0887}} \color[HTML]{F1F1F1} \bfseries 26.407 & {\cellcolor[HTML]{0D0887}} \color[HTML]{F1F1F1} \bfseries 6.352 & {\cellcolor[HTML]{6900A8}} \color[HTML]{F1F1F1} 846.752 & {\cellcolor[HTML]{B12A90}} \color[HTML]{F1F1F1} 214.710 \\
            Mix.Ind & {\cellcolor[HTML]{7701A8}} \color[HTML]{F1F1F1} 16.035 & {\cellcolor[HTML]{D9586A}} \color[HTML]{F1F1F1} 26.543 & {\cellcolor[HTML]{0D0887}} \color[HTML]{F1F1F1} 6.355 & {\cellcolor[HTML]{F0F921}} \color[HTML]{000000} 1028.382 & {\cellcolor[HTML]{F0F921}} \color[HTML]{000000} 258.125 \\
            Mix.CD & {\cellcolor[HTML]{7401A8}} \color[HTML]{F1F1F1} 16.032 & {\cellcolor[HTML]{E56B5D}} \color[HTML]{F1F1F1} 26.558 & {\cellcolor[HTML]{F0F724}} \color[HTML]{000000} 7.462 & {\cellcolor[HTML]{FDB52E}} \color[HTML]{000000} 992.077 & {\cellcolor[HTML]{FCA338}} \color[HTML]{000000} 243.131 \\
            Mix.TD & {\cellcolor[HTML]{6C00A8}} \color[HTML]{F1F1F1} 16.024 & {\cellcolor[HTML]{BA3388}} \color[HTML]{F1F1F1} 26.511 & {\cellcolor[HTML]{FEC029}} \color[HTML]{000000} 7.321 & {\cellcolor[HTML]{F3F027}} \color[HTML]{000000} 1023.270 & {\cellcolor[HTML]{F2F227}} \color[HTML]{000000} 256.778 \\
        \bottomrule
    \end{tabular}
    \caption{Error statistics. All error metrics are averaged over $t$, $h$ and $d$. The lowest value is \textbf{highlighted}. Background color corresponds to forecasting accuracy. $\text{ES}_{3\text{H}}$ denotes the energy score for the last 3 hours of trading.}
    \label{tab:errors_scoring_rules}
\end{table}

Remember that we have both, the naive model and the mixture model in in at least two versions: one version assuming independence between different delivery hours and at least one version that considers the correlation structure. Additionally, the \textbf{RW.Emp} considers the correlation structure implicitly. The results for the energy score show that considering the correlation structure leads to (significantly, see Figure \ref{fig:errors_dmtest}) better forecasts than assuming independence if the remaining model structure is unchanged. This holds for moving from the \textbf{Naive.Ind} to the \textbf{Naive.Dep} and for moving from the \textbf{Mix.Ind} to \textbf{Mix.CD}. Interestingly, modelling the dependence structure in a potentially time-varying fashion does not improve the forecasting performance. For the energy score, the difference in ordering for the models between the full path and the last three hours of trading. Note however, that the scale of both is not directly comparable. This is an important result for modelling the intraday market in applications such as battery/storage optimisation, where the neglecting the correlation structure can therefore lead to too optimistic results \cite{nolzen2022market, lohndorf2023value}. 

On an aggregate level, we see that the \textbf{RW.Emp} yields the best point forecasting performance (MAE and RMSE) and the \textbf{Naive.Dep} yields the best probabilistic forecasting performance. The Diebold-Mariano test confirms the statistical significance of superior probabilistic forecasting forecasts for the \textbf{Naive.Dep}. We note that the forecasting performance of the mixture models is not as good as the rather simple benchmark models. Additionally, we note that the \textbf{Mix.Ind} exhibits a lower CRPS than the \textbf{Mix.CD} and \textbf{Mix.TD}, which suggests some some cross-propagation of errors. On the other hand, the \textbf{Mix.Ind} yields worse scores for the energy score than its sister models including a dependence structure.

\begin{figure}
    \centering
    \includegraphics[width=\textwidth]{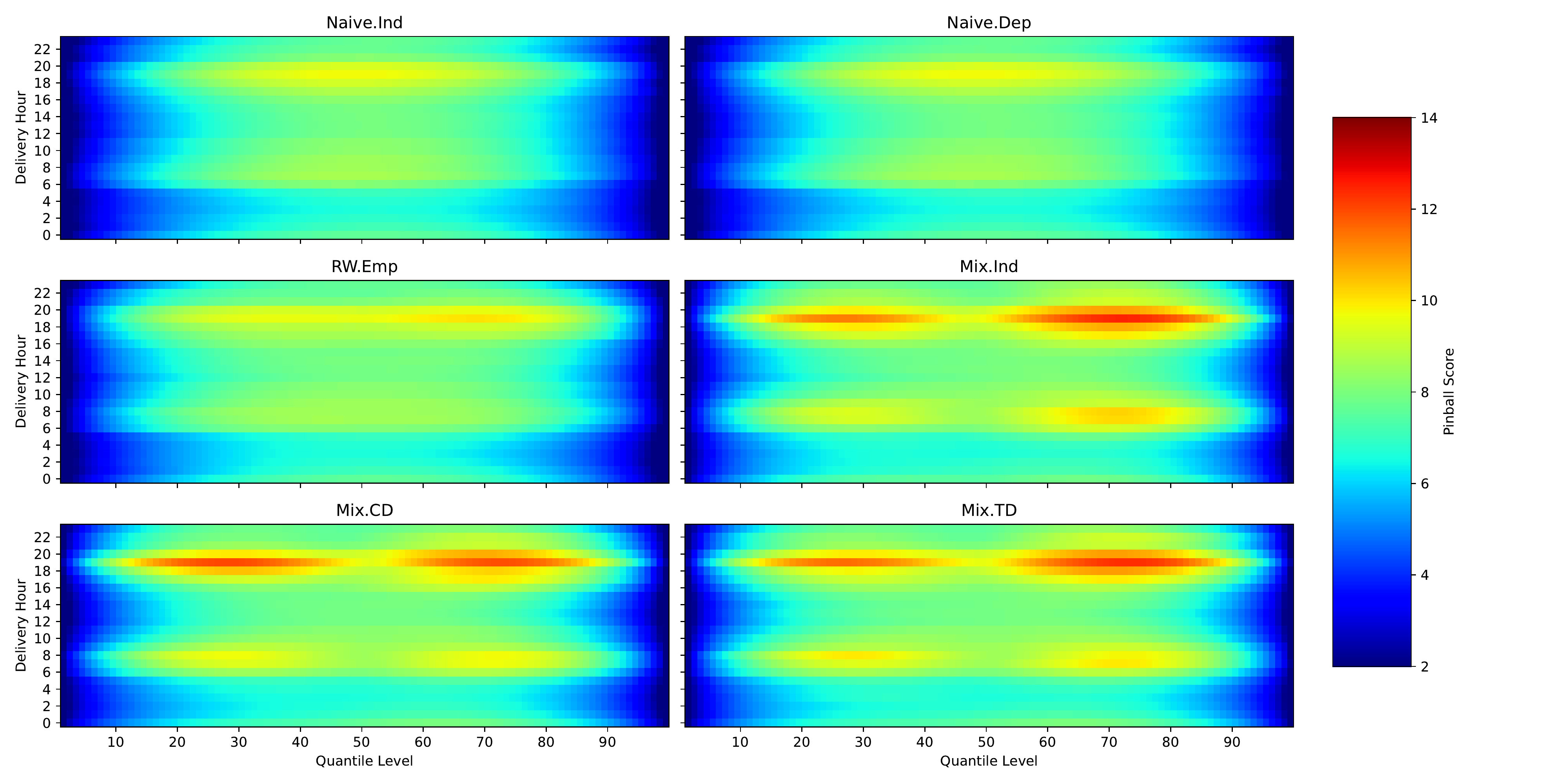}
    \caption{Pinball Score by quantile level $\tau$ and delivery hour $h$.}
    \label{fig:errors_pinball_quantile}
\end{figure}

The hourly shape of forecasts errors throughout the day is depicted in Figure \ref{fig:errors_by_hour}. It follows the typical shape of prices in electricity prices, we see low forecast errors in the morning and higher errors through the day. Let us note that the benchmark models exhibit a slightly higher MAE during the afternoon peak, but show lower RMSE and ES during the same periods. Figure \ref{fig:errors_pinball_quantile} presents the PB across the quantile range and the delivery hour. We can see that the highest errors occur during the afternoon peak hours and in the higher distribution quantiles. Note that the pinball loss only measures the marginal fit to the distribution. In an analysis of the in-sample, transformed observations we also note that throughout the rolling window study, the calibration decreases and we experience an underdispersed forecast (see Figure \ref{fig:diagnostics_insample_fit} in the Appendix). 

Our results also emphasize the importance of robust approaches to modelling and forecasting in periods of high volatility and black swan events such as the Russian invasion of Ukraine and the subsequent energy crisis in Europe 2022/23. The \textbf{Naive.Dep} and \textbf{Naive.Ind} model already have shown very good probabilistic forecasting performance in the respective studies of \cite{hirsch2022simulation, narajewski2020ensemble} and are robust to extreme events, as out-of-support situations with extreme prices cannot happen for these models. Additionally, our results can be viewed in the light of the market efficiency hypothesis. Our result indicate that including more data, especially data that does not change throughout the trading period (such as day-ahead forecasts) does not improve the modelling or the price. The superior performance of the benchmark models, especially of the \textbf{RW.Emp}, which ensures the Martingale assumption, underscores this notion. Similar results with respect to market efficiency have been found by \cite{narajewski2020econometric}. 

\section{Conclusion} \label{sec:discussion}

This paper presents a a forecasting study for multivariate, simulation-based forecasting for intraday electricity markets. We provide insight in the dependence structure of short-term electricity markets and extend previous works of \cite{narajewski2020ensemble, hirsch2022simulation} to include cross-product price effects.

We develop a probabilistic model for the marginal distribution of the intraday price path, accounting for the impact of fundamental driving variables on the location, scale and shape parameters. As novelty, we employ copulas to model the (time-dependent) dependency between different delivery periods and the according parallel trading sessions and allow the dependency parameter to be time-dependent. We validate our results in a forecasting study for the German intraday electricity market. Our results indicate that modelling the dependence structure between the different trading session improves the forecasting performance. Additionally, we provide evidence on market efficiency in the German short-term market. Our fundamental and parametric approach allows us to shed further light on the impact of the cross-border shared order books of the single intraday coupling (SIDC) and the driving factors of the distribution parameters. Our case study employs data from the German intraday electricity market, but our method is directly transferable to other European electricity markets.

While we are able to show that modelling the dependence structure improves the forecasting performance, our methods can surely be improved and our results offer a multitude of further research areas: First, a further investigation of the dependency structure seems worthwhile. However, the improved modelling of the correlation structure is dependent on having a suitable probabilistic marginal model at hand. We note that the proposed mixture of Johnson's $S_U$ distribution still does not provide a perfect fit and struggles to cope with periods of extreme volatility. Hence, the modelling of the price distribution is an important field for further research. Third, we provide new evidence on the impact of the single intraday coupling (SIDC) on the price distribution and trading activity. As the SIDC system is dynamically changing and little researched, this might be an interesting direction for further research into the micro-structure of intraday electricity markets. The literature on intraday markets is still scarce compared to the fast growth of renewable energy sources and intraday electricity markets.

\bibliographystyle{abbrv}
\bibliography{references}

\appendix

\begin{figure}
    \centering
    \includegraphics[width=0.8\textwidth]{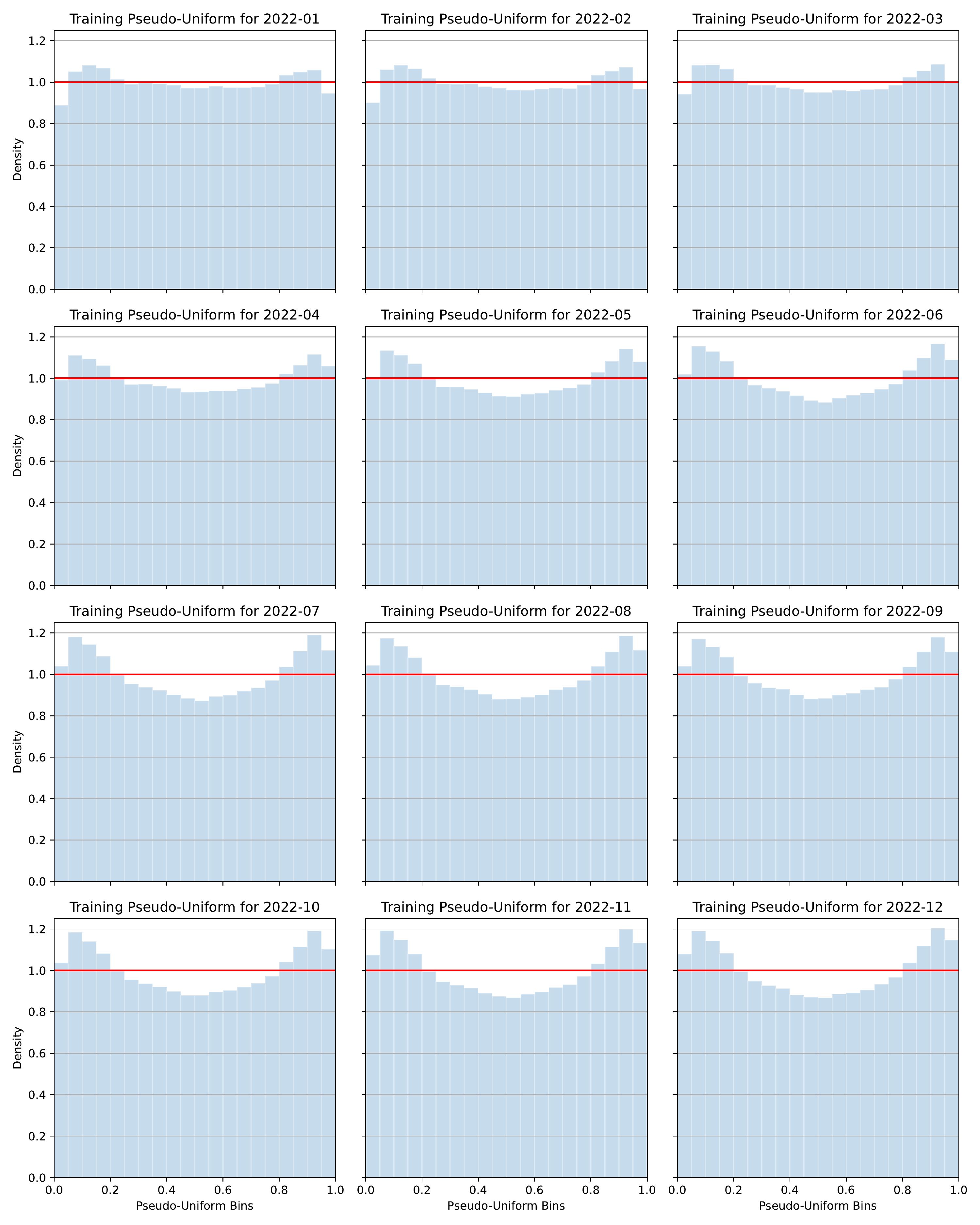}
    \caption{Diagnostic plots for the in-sample pseudo-uniform observations after each refit during the forecasting study.}
    \label{fig:diagnostics_insample_fit}
\end{figure}

\begin{figure}
    \centering
    \includegraphics[width=0.8\textwidth]{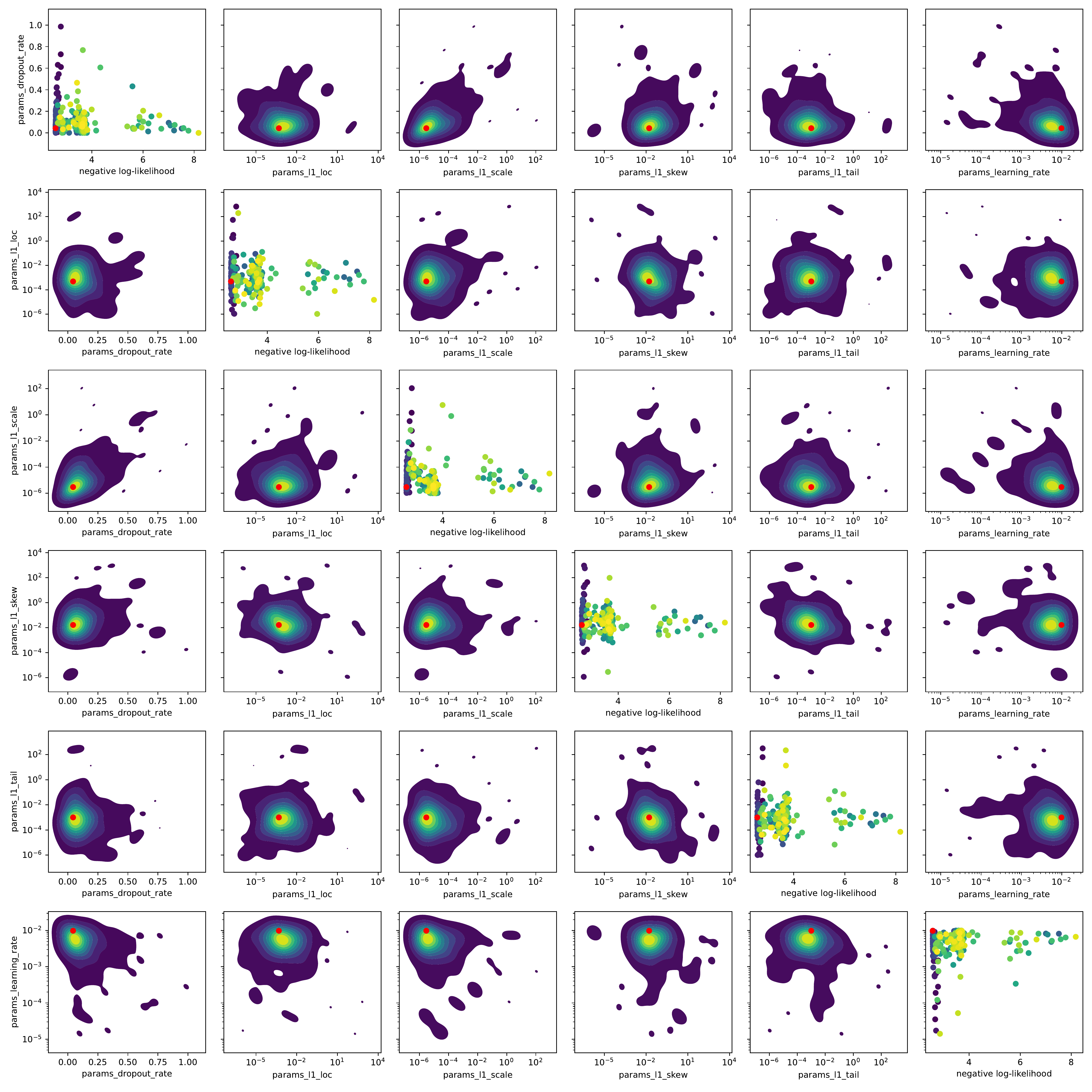}
    \caption{Diagnostic plots for our hyperparameter tuning using the \texttt{optuna} framework. We show the bivariate distribution of the explored hyperparameters and the scatter plot of each hyperparameter towards the  negative log-likelihood on the diagonal. Color represents increasing the trial number from dark to yellow. The red dot represents the best trial.}
    \label{fig:diagnostics_optuna}
\end{figure}

\end{document}

%% file: figures/fig_market_structure_shortterm.tex
\begin{tikzpicture}
\begin{footnotesize}

\coordinate (A) at (0,0);
\coordinate (B) at (14, 0); 

\coordinate[label={[above, align=center, font=\scriptsize] Day-ahead \\ Spot Auction}] (11) at (0.5,2);
\coordinate[label={[below, align=center] $d-1$, \\ 12.00 }] (12) at (0.5,-0.5);

\coordinate[label={[above, align=center, font=\scriptsize] Intraday \\ Market opens}] (21) at (3,2);
\coordinate[label={[below, align=center] $d-1$, \\ 15.00 }] (22) at (3,-0.5);

\coordinate[label={[above, align=center, font=\scriptsize] 1st Wave \\ SIDC opens}] (31) at (5,2);
\coordinate[label={[below, align=center] $d-1$, \\ 18.00 }] (32) at (5,-0.5);

\coordinate[label={[above, align=center, font=\scriptsize] 2nd Wave \\ SIDC opens}] (51) at (7,2);
\coordinate[label={[below, align=center] $d-1$, \\ 22.00 }] (52) at (7,-0.5);

\coordinate[label={[above, align=center, font=\scriptsize] SIDC \\ closes}] (61) at (10.5,2);
\coordinate[label={[below, align=center] $d$, \\ $h-60$ \\ min}] (62) at (10.5,-0.5);

\coordinate[label={[above, align=center, font=\scriptsize] Market \\ closes}] (71) at (11.5,1.25);
\coordinate[label={[below, align=center] $d$, \\ $h-30$ \\ min}] (72) at (11.5,-0.5);

\coordinate[label={[above, align=center, font=\scriptsize] Control zones \\ close}] (81) at (12.5,0.5);
\coordinate[label={[below, align=center] $d$, \\ $h-5$ \\ min}] (82) at (12.5,-0.5);

\coordinate[label={[above, align=center, font=\scriptsize] Delivery}] (91) at (13.5,2);
\coordinate[label={[below, align=center] $d$, $h$}] (92) at (13.5,-0.5);

\draw [->,line width=1.25pt] (A) -- (B);
\draw [-] (11) -- (12);
\draw [-] (21) -- (22);
\draw [-] (31) -- (32);
\draw [-] (51) -- (52);
\draw [-] (61) -- (62);
\draw [-] (71) -- (72);
\draw [-] (81) -- (82);
\draw [-,line width=1.25pt] (91) -- (92);

\end{footnotesize}
\end{tikzpicture}